**Impact of hydrogen addition, up to 20 % (mol/mol), on the thermodynamic ($p$, $\rho$, $T$) properties of a reference high-calorific natural gas mixture with significant ethane and propane content**

Daniel Lozano-Martín[1], Heinrich Kipphardt[2], Peyman Khanipour[2], Dirk Tuma[2], Alfonso Horrillo[3,4], César Chamorro[4,*]

[1] GETEF, Departamento de Física Aplicada, Facultad de Ciencias, Universidad de Valladolid, Paseo de Belén, 7, E-47011 Valladolid, Spain.

[2] BAM Bundesanstalt für Materialforschung und -prüfung, D-12200 Berlin, Germany.

[3] CIDAUT Fundación, Plaza Vicente Aleixandre Campos, 2, E-47151, Boecillo-Valladolid, Spain

[4] MYER, Departamento de Ingeniería Energética y Fluidomecánica, Escuela de Ingenierías Industriales, Universidad de Valladolid, Paseo del Cauce, 59, E-47011 Valladolid, Spain.

## Abstract

Injecting hydrogen into the natural gas grid supports gradual decarbonization. To check the accuracy of equations of state for hydrogen-enriched natural gas mixtures, precise density data from well-characterized reference mixtures are essential. In a prior study, we provided experimental measurements for a natural gas constituted mainly of methane and for two derived hydrogen-enriched mixtures. In the present study, being the second and final part of our investigation, density measurements for a high-calorific natural gas with significant ethane and propane content, along with two hydrogen-enriched variants (10 and 20 mol-% hydrogen) are provided. The mixtures are gravimetrically prepared following ISO 6142-1. Density measurements, conducted with a single-sinker densimeter at temperatures from 260 to 350 K and pressures up to 20 MPa, are compared with three equations of state: AGA8-DC92, GERG-2008, and an improved GERG-2008. Results indicate that all models perform better for methane-dominant mixtures than for those containing heavier hydrocarbons.

* Corresponding author e-mail: cesar.chamorro@uva.es.

List of Abbreviations

| | |
|---|---|
| *Nomenclature* | |
| $k$ | Coverage factor |
| $m$ | Mass, kg |
| $M$ | Molar mass, kg/mol |
| N | Number of components of a mixture |
| $p$ | Pressure, MPa |
| $T$ | Temperature, K |
| $u$ | Standard uncertainty |
| $U$ | Expanded uncertainty |
| $V$ | Volume, $m^3$ |
| $W$ | Balance reading, kg |
| $W_s$ | Wobbe index |
| $x$ | Molar fraction |
| *Abbreviations* | |
| *AARD* | Average absolute value of relative deviation |
| AGA | American Gas Association |
| AGA8-DC92 | AGA8-DC92 equation of state for mixtures |
| BAM | Federal Institute for Materials Research and Testing |
| *BiasRD* | Average relative deviation |
| GERG | Groupe Européen de Recherches Gazières |
| GERG-2008 | GERG-2008 equation of state for mixtures |
| EoS | Equation of State |
| GC | Gas chromatography |
| GUM | Guide to the Expression of Uncertainty in Measurement |
| H2NG | Hydrogen-enriched natural gas |
| HHV | Higher heating value |
| LNG | Liquefied natural gas |
| *MaxRD* | Maximum relative deviation |
| NG | Natural gas |
| *RMSRD* | Root mean square relative deviation |
| UVa | University of Valladolid |
| VLE | Vapor-liquid equilibria |
| *Greek symbols* | |
| $\alpha$ | Calibration factor |
| $\varepsilon_\rho$ | Apparatus-specific constant |
| $\phi_0$ | Coupling factor |
| $\kappa_T$ | Isothermal compressibility, $Pa^{-1}$ |
| $\rho$ | Density, $kg \cdot m^{-3}$ |
| $\rho_0$ | Reducing density, $kg \cdot m^{-3}$ |
| $\rho_n$ | Normalized density |
| $\chi_s$ | Specific magnetic susceptibility, $m^3 \cdot kg^{-1}$ |
| $\chi_{s0}$ | Specific reducing magnetic susceptibility, $m^3 \cdot kg^{-1}$ |
| *Subscripts* | |
| AGA8-DC92 | Calculated from AGA8-DC92 equation of state |
| EoS | Calculated from an equation of state |
| exp | Experimental |
| fluid | fluid |
| GERG-2008 | Calculated from GERG-2008 equation of state |
| GERG-imp | Calculated from enhanced version of the GERG-2008 equation of state |
| i | i-th component |
| j | j-th component |
| MP | Measuring position |
| r | Relative |
| s | Sinker (however, “specific” in case of $\chi_s$ and $\chi_{s0}$) |
| T | Overall |
| Ti | Titanium |
| Ta | Tantalum |
| ZP | Zero position |

## 1. Introduction

Hydrogen-enriched natural gas (H2NG) is a blend of natural gas (NG) and hydrogen that is compatible with current NG infrastructure. Injecting hydrogen directly into the NG grid provides an economical solution for transporting and storing hydrogen generated from intermittent renewable energy sources. Furthermore, injecting hydrogen from renewable sources, along with the partial or total substitution of NG with biomethane and/or synthetic methane (e-methane), is seen as a sustainable approach for the progressive decarbonization of heating and power systems [1–3]. While hydrogen can theoretically be mixed with NG, biomethane, or e-methane, in any ratio, mixtures containing up to 20 % hydrogen by volume are considered the most feasible near-term option [4–8].

Several challenges must be resolved before hydrogen can be injected into the NG grid on a large scale [9]. From the perspective of end-use appliances, the combustion characteristics of the gas mixture may change, potentially affecting equipment performance [10–13]. Another critical concern is hydrogen embrittlement, as it can compromise the mechanical integrity of iron and steel pipelines [14–16]. Additionally, the design of systems for the production, transport, and storage of H2NG mixtures, particularly for custody transfer applications, relies heavily on precise and consolidated data for volumetric and calorific properties which usually are derived from reference equations of state (EoS) [17,18].

The AGA8-DC92 [19] and the GERG-2008 [20,21] EoS are established among both industry and academia as reference EoS for NG mixtures. The underlying framework of these EoS is built on experimental data of the constituting pure components and of the consequent binary mixtures. Experimental data on pure NG components and binary mixtures have been essential for developing these EoS and remain crucial for their refinement. [22], especially for binary mixtures of the main components of NG with hydrogen [23–28]. Additionally, high-accuracy experimental thermophysical data from multicomponent H2NG mixtures—prepared with minimal composition uncertainty—are especially valuable for testing the reliability of these equations originally developed for NG mixtures when applied to H2NG mixtures [29–31].

In a recent paper [32], our group evaluated the performance of these reference EoS in predicting the thermophysical properties of a high-calorific NG (H2-free) mixture composed mainly of methane (> 97 %) and of two derived H2NG mixtures with nominal hydrogen concentrations of 10 % and 20% (mol/mol). We obtained density measurements for these three mixtures at temperatures ranging from (250 to 350) K and pressures up to 20 MPa, using a high-precision single-sinker magnetic suspension densimeter. Overall, the experimental density values for the $H_2$-free NG mixture aligned well with the predicted values, within the stated uncertainty of the equations. However, for the H2NG mixtures, larger deviations from the predicted densities were observed, particularly at lower temperatures and higher pressures. The maximum relative deviations reached nearly 0.2 %—exceeding the claimed uncertainty of 0.1 % for all three EoS.

In this follow-up work, in a similar way, we offer experimental density data for a new standard high-calorific NG mixture and for two derived H2NG mixtures, also with nominal hydrogen concentration of 10 % and 20 % (mol/mol). In this case, the corresponding initial NG mixture deviates from the initial mixture of the first study [32], with significant amounts of ethane (9 % (mol/mol) compared to 0.4 %) and propane (3 % (mol/mol) compared to 0.2 %), while the content of methane is reduce to 85 % (mol/mol). All investigated mixtures were gravimetrically prepared following ISO 6142-1 [33] to ensure reference quality. The density measurements were conducted using the same high-precision single-sinker densimeter as in the previous work, covering a temperature range from (260 to 350) K and pressures up to 20 MPa. The experimental density results obtained in this study are compared with the aforementioned three reference EoS for NG related mixtures: the AGA8-DC92, the GERG-2008, and an enhanced version of the GERG, referred to in this study as “improved-GERG-2008”.

## 2. Experimental

### 2.1. Mixture preparation

Three synthetic NG mixtures were made at the Federal Institute for Materials Research and Testing (BAM, Bundesanstalt für Materialforschung und -prüfung) in Berlin, Germany. These mixtures were labeled G 432 (the $H_2$-free NG mixture, cylinder no.: 2031-200915), G 455 (the hydrogen-enriched derivative, G 432 + 10 % $H_2$, 6079-210111), and G 456 (hydrogen-enriched, G 432 + 20 % $H_2$, 2059-210201). The first mixture, G 432, is an 11-components, $H_2$-free NG, with methane (85 %) being the matrix component, but with significant amounts of ethane (9 %), propane (3 %), carbon dioxide (1.4 %), and nitrogen (1 %). The derived mixtures, G 455 and G 456, were created by adding hydrogen to G 432, achieving nominal hydrogen concentrations of 10 % and 20 % (mol/mol), respectively.

Each mixture was prepared in 10 $dm^3$ aluminum cylinders using a gravimetric method in accordance with ISO 6142-1 [33], a procedure for the preparation of reference mixtures ensuring minimal uncertainty in composition [32]. The pure components, listed in Table S1, were used directly without further purification to create various premixtures and dilutions in several filling stages. Detailed filling procedures are given in the Supplement. The mass of each gas component was measured using either an electronic comparator balance (Sartorius LA 34000P-0CE) or a high-precision mechanical balance (Voland HCE 25). To validate each target mixture (G 432, G 455, and G 456), two calibration mixtures were independently prepared following BAM's accredited certification protocol, ensuring no correlation between the sample and calibration mixtures. After finishing gravimetric preparation, the gas mixtures were thoroughly homogenized through heating and rolling. The molar compositions ($x_i$) and their resultant expanded ($k = 2$) uncertainties in absolute terms, $U(x_i)$, are provided in Table 1.

Before shipment to the University of Valladolid (UVa), each mixture underwent composition validation by Gas Chromatography (GC) at BAM, using a multichannel process analyzer (Siemens MAXUM II, Siemens AG). The validation protocol followed the "bracketing" procedure outlined in the ISO 12963 standard [34]. The results are given in Table S3. Further details of this validation method can be found in a previously published paper [31]. The uncertainties in the concentration values of each component in both the studied mixtures and their validation mixtures were determined using the law of propagation of uncertainty, following the guidelines outlined in the Guide to the Expression of Uncertainty in Measurement (GUM) [35].

The molar mass $M$, normalized density $\rho_0$, higher heating value $HHV$, and Wobbe index $W_s$ for the three gas mixtures, at reference conditions of 288.15 K and 0.101325 MPa, were calculated using the GERG-2008 EoS, included in the REFPROP 10 [36,37] software, based on their normalized compositions. These values are presented in Table 1 as complementary information. The results show that the G 432 mixture is a high-calorific NG blend. When hydrogen is introduced into the mixture, there is a noticeable diminution in normalized density, higher heating value, and Wobbe index, but all variables still remain within the limits to categorize these mixtures as high-calorific NG blends. Specifically, when 10 % hydrogen (G 455) or 20 % hydrogen (G 456) is added to the NG, the higher heating value per unit volume decreases by 7 % and 14 %, respectively, compared to the original G 432 mixture with no hydrogen. The Wobbe index shows smaller changes, decreasing by 2.7 % and 5.4 %, respectively, for G 455 and G 456.

For the critical properties, calculated also by the GERG-2008 Eos by using the Refprop software, the critical point of the $H_2$-free NG mixture (G 432) occurs at a temperature of 222.4 K and a pressure of 7.2 MPa, with the cricondentherm at 257.4 K and 4.8 MPa, and the cricondenbar at 238.8 K and 7.9 MPa. In the case of the G 455 mixture (G 432 with 10% $H_2$), the critical point shifts to 215.7 K and 10.0 MPa, the cricondentherm to 257.0 K and 5.4 MPa, and the cricondenbar to 224.1 K and 10.1 MPa. For the G 456 mixture (G 432 with 20% $H_2$), the critical point is at 217.1 K and 13.6 MPa, the cricondentherm at 256.4 K and 6.1 MPa, and the cricondenbar cannot be computed. It is worth mentioning that neopentane was not incorporated into the mixture models used in this study; consequently, its contribution was accounted for by adding it to the n-pentane concentration, as recommended by the GERG-2008 Eos.

## 2.2. Description of the equipment setup and measurement procedure

The experimental procedure in this work was carried out using a single-sinker magnetic suspension densimeter, widely recognized as one of the most accurate methods for determining fluid density across a broad range of temperatures and pressures, and following the same protocol as in the previous study [32]. This setup includes a pressurized diamagnetic cell, within which a monocrystalline silicon sinker with a precisely calibrated volume ($V$s = 226.4440 ± 0.0026 cm$^3$) is suspended and surrounded by the sample gas. The buoyant force acting on the sinker is transmitted to a highly sensitive microbalance (XPE205DR, Mettler Toledo GmbH), which is positioned above the cell at ambient pressure and temperature, via a magnetic coupling mechanism. The underlying measurement principles were initially developed by the group of Wagner at the University of Bochum, Germany [38–41], who first implemented them with two-sinker systems with extremely high accuracy, particularly at low densities, by compensating for adsorption effects. This was later implemented to single-sinker systems, which, while simpler, provide equally precise measurements at higher densities [42,43].

The equation used to estimate the density of the fluid is:

$$\rho_{fluid} = \frac{\phi_0 m_s + (m_{\mathrm{Ti}} - m_{\mathrm{Ta}}) + (W_{ZP} - W_{MP})/\alpha}{V_s(T,p)} \frac{1}{\phi_0} + \frac{\varepsilon_\rho}{\phi_0} \frac{\chi_s}{\chi_{s0}} \left( \frac{\rho_s}{\rho_0} - \frac{\rho_{fluid}}{\rho_0} \right) \rho_{fluid} \quad (1)$$

where the terms $m$, $V$, and $\rho$ denote the mass, volume, and density, respectively, while the subscripts fluid, and s stand for the fluid, and sinker (however, "specific" in case of $\chi_s$ and $\chi_{s0}$). The subscripts Ti, and Ta refers to titanium and tantalum counterweights, while ZP, and MP refers to zero and measuring positions of the magnetic coupling. The calibration factor $\alpha$ is determined by weighing two calibrated counterweights of tantalum and titanium, which are alternately placed in the upper pan of the microbalance using an automatic changing device. The counterweights have nearly identical volumes, with their mass difference approximating that of the sinker.

The measurement process is performed by subtracting two separate balance readings. In the zero position (ZP) the electromagnet hanging from the lower hook of the balance pulls the sinker support without raising the sinker, whereas in the measuring position (MP) a greater force is applied to the permanent magnet, fixed to the upper end of the sinker support inside the cell, lifting the silicon sinker. The difference in the balance readings, $W$, between these two positions effectively cancels out the weights of the sinker support, magnets, and balance hook.

Due to small differences in the vertical positions of the ZP and MP, along with potential instabilities in magnet alignment, the density measurement requires correction for the so-called *force transmission error*. This error consists of two components, namely the apparatus-specific effect and the fluid-specific effect. The apparatus-specific effect is represented by $\phi_0$ in Equation (1) and is determined by calculating the sinker weight in vacuum once all the data for an isotherm has been collected. This correction must always be applied to avoid significant errors [44]. The fluid-specific effect, given by the second term on the right-hand side of Equation (1), depends on (a) the specific magnetic susceptibility of the fluid $\chi_s$, (b) the apparatus-specific constant $\varepsilon_\rho$, and (c) the reducing constants $\chi_{s0} = 10^{-8}$ m$^3$·kg$^{-1}$ and $\rho_0$ = 1000 kg·m$^{-3}$. The value of $\varepsilon_\rho$ was previously determined for our densimeter, along with its dependence on temperature and density, and results were presented in a previous work [45]. Unlike the apparatus-specific effect, the fluid-specific effect is minimal for diamagnetic fluids but can build up a 3% error for paramagnetic fluids, where $\chi_s$ can be up to 100 times greater and temperature-dependent [46–48].

The fluid pressure is measured using two quartz crystal transducers: one for low pressures (0–3 MPa) (Digiquartz 2300A-101) and the other for higher pressures (3–20 MPa) (Digiquartz 43KR-HHT-101), both from Paroscientific Inc. The expanded ($k$ = 2) uncertainty for the low-pressure transducer is $U(p) = (6.0\cdot10^{-5}(p/\text{MPa}) + 2\cdot10^{-3})$ MPa, and for the high-pressure transducer, $U(p) = (7.5\cdot10^{-5}(p/\text{MPa}) + 4\cdot10^{-3})$ MPa.

The cell temperature is controlled by an oil thermal bath (Dyneo DD-1000F, Julabo GmbH) and an electrical heating cylinder with a temperature controller (MC-E, Julabo GmbH). Temperature is measured, with an expanded ($k$ = 2) uncertainty $U(T)$ = 0.015 K, by means of two platinum resistance thermometers (SPRT-25, Minco Products Inc.) and an AC resistance bridge (ASL F700, Automatic Systems Laboratory).

A more comprehensive explanation of the equipment and experimental technique is available in earlier publications [49,50].

## 3. Experimental results

### 3.1. Measurement uncertainty analysis

The experimental overall expanded ($k$ = 2) uncertainty $U_T(\rho_{exp})$ for the density measurements is presented in Table 2, shown in both absolute and relative terms. This uncertainty accounts for contributions from the density determination uncertainty, $U(\rho_{exp})$, which has been comprehensively assessed in previous studies [45,50] for our single-sinker densimeter as a function of both density and specific magnetic susceptibility:

$$U(\rho_{\text{exp}})/(\text{kg}\cdot\text{m}^{-3}) = 2.5\cdot10^{4}\cdot\chi_s/(\text{m}^3\cdot\text{kg}^{-1}) + 1.1\cdot10^{-4}\cdot\rho_{\text{exp}}/(\text{kg}\cdot\text{m}^{-3}) + 2.3\cdot10^{-2} \quad (2)$$

combined with the uncertainties from pressure, $u(p)$, temperature, $u(T)$, and composition, $u(x_i)$, following the law of uncertainty propagation [51]:

$$U_{\text{T}}(\rho_{\text{exp}}) = 2\left[u(\rho_{\text{exp}})^2 + \left(\left.\frac{\partial\rho}{\partial p}\right|_{T,x} u(p)\right)^2 + \left(\left.\frac{\partial\rho}{\partial T}\right|_{p,x} u(T)\right)^2 + \sum_i\left(\left.\frac{\partial\rho}{\partial x_i}\right|_{T,p,x_j\neq x_i} u(x_i)\right)^2\right]^{0.5} \quad (3)$$

The partial derivatives of the mixture density with respect to pressure and temperature are estimated using REFPROP 10 software [36,37], which applies the improved GERG-2008 EoS [52–54]. Here, the most significant contribution to uncertainty arises from $U(p)$ and $U(\rho_{exp})$, reaching values as high as 0.09 kg·m$^{-3}$ (0.44 %). Contributions from $U(x_i)$ and $U(T)$, are much smaller, below 0.017 kg·m$^{-3}$ (0.01 %) and 0.0028 kg·m$^{-3}$ (0.002 %), respectively. The overall experimental expanded ($k$ = 2) uncertainty for the three mixtures varies from (0.031 to 0.096) kg·m$^{-3}$, equivalent to 0.033 % to 0.57 %.

### 3.2. Density measurements

Tables 3, 4, and 5 provide the experimental ($p$, $\rho$, $T$) data for mixtures G 432 ($H_2$-free high calorific NG), G 455 (G 432 + 10 % $H_2$), and G 456 (G 432 + 20 % $H_2$), the expanded ($k$ = 2) uncertainty in density for these measurements calculated using Equation (3), in both absolute and percentage terms, and the relative deviations of the experimental densities from those predicted by the AGA8-DC92 EoS, GERG-2008 EoS, and improved GERG-2008 EoS. Measurements were recorded at five temperatures—260 K, 275 K, 300 K, 325 K, and 350 K—with pressure successively decreasing in 1 MPa steps from 20 MPa down to 1 MPa. Figure 1 presents the data points for the three mixtures, along with the saturation curves calculated from the improved GERG-2008 EoS, the typical operational ranges for pipeline conditions in the gas industry, and the approved application limits for both the AGA8-DC92 and GERG-2008 EoS models.

The AGA8-DC92 EoS [19], developed by the American Gas Association, is a reference model for NG mixtures. Originally formulated as a virial expansion of the compressibility factor, it was later adapted into an explicit Helmholtz energy form to address both calorific and volumetric properties [55]. Its application range covers gas and supercritical phases within (250 to 350) K and up to 30 MPa.

The GERG-2008 EoS [20,21], a model from the Groupe Européen de Recherches Gazières and widely used in Europe, expands the scope of AGA8-DC92 to include liquid and vapor-liquid equilibria (VLE) over a broader range of (60 to 700) K and pressures up to 70 MPa [52]. Both models are capable to describe thermophysical properties of NG mixtures with up to 21 components at pipeline conditions. Their accuracy depends on implemented available experimental data for VLE, density, speed of sound, enthalpy, and heat capacity. To enhance the applicability of GERG-2008 EoS, two major updates were recently made. The first modification targets LNG applications by improving accuracy in the subcooled liquid region from (90 to 180) K up to 10 MPa, using new and reparametrized departure functions for binary mixtures with methane [53]. This adjustment improved density and calorific property predictions for LNG mixtures [56,57]. The second modification specifically focuses on hydrogen-rich mixtures, critical for hydrogen economy processes, by using newer, updated pure-fluid equations and refining binary-specific departure functions [54]. This new approach is able to fix phase envelope issues observed with the original GERG-2008 EoS at low temperatures and reduced deviations in VLE, density, and speed of sound data. Together, these modifications form what we refer to in this study as the "improved GERG-2008" EoS.

## 4. Discussion

### 4.1. Relative deviations of experimental data from the reference equations of state

Figures 2, 3, and 4 display the percentage relative deviations of experimental density data from the values calculated using the AGA8-DC92 EoS, GERG-2008 EoS, and improved GERG-2008 EoS models for the G 432, G 455, and G 456 mixtures, respectively. Table 6 presents a statistical comparison of the experimental density data obtained in this study relative to the three EoS models applied. Table 6 also includes the results of the statistical comparison for the three mixtures of our previous work [32]. In Table 6, the statistical indicators are defined as follows: *AARD* (average absolute value of relative deviations), *BiasRD* (average relative deviation), *RMSRD* (root mean square relative deviation), and *MaxRD* (maximum relative deviation), as given by Eqs. (4) to (7):

$$\mathrm{AARD} = \frac{1}{N}\sum_{i=1}^{N}\left|10^2\,\frac{\rho_{\mathrm{i,exp}}-\rho_{\mathrm{i,EoS}}}{\rho_{\mathrm{i,EoS}}}\right| \tag{4}$$

$$\mathrm{BiasRD} = \frac{1}{N}\sum_{i=1}^{N}\left(10^2\,\frac{\rho_{\mathrm{i,exp}}-\rho_{\mathrm{i,EoS}}}{\rho_{\mathrm{i,EoS}}}\right) \tag{5}$$

$$\mathrm{RMSRD} = \sqrt{\frac{1}{N}\sum_{i=1}^{N}\left(10^2\,\frac{\rho_{\mathrm{i,exp}}-\rho_{\mathrm{i,EoS}}}{\rho_{\mathrm{i,EoS}}}\right)^2} \tag{6}$$

$$\mathrm{MaxRD} = \max\left|10^2\,\frac{\rho_{\mathrm{i,exp}}-\rho_{\mathrm{i,EoS}}}{\rho_{\mathrm{i,EoS}}}\right| \tag{7}$$

The density values given by the different EoS were calculated using the normalized composition without impurities given in Table 1. The effect of impurities in the density calculated by the different EoS was estimated to be far below the experimental uncertainty and the deviations.

Relative deviations between the experimental density data for the $H_2$-free NG mixture (G 432) and the AGA8-DC92 EoS, as shown in Figure 2(a), generally fall within the claimed uncertainty of this EoS ($U(\rho_{\mathrm{EoS}})$ = 0.1 %), except at the lower temperature of 260 K and pressures between 2 and 10 MPa, where deviations can reach up to -0.15 %. For the same temperature, but at pressures above 9 MPa and up to 16 MPa, deviations between experimental density data and the GERG-2008 EoS increase, reaching -0.21 % (see Figure 2(b)), and are even larger (-0.28%) when compared to the improved GERG-2008 (see Figure 2(c)). For the latter EoS, even at 275 K and pressures above 11 MPa, the density data show deviations exceeding the stated uncertainty, reaching -0.15 %. Notably, this hydrogen-free NG mixture (G 432) contains substantial amounts of ethane and propane. By contrast, results from our previous study [32] on a hydrogen-free NG mixture composed primarily of methane (> 97 %) showed that all three EoS models matched the experimental density data accurately, with maximum relative deviations around 0.05 %.

For the G 455 $H_2$-enriched NG mixture (G 432 + 10 % $H_2$), the relative deviations from the three EoS models are shown in Figure 3. In this case, deviations from the AGA8-DC92 EoS (Figure 3(a)) remain within the stated uncertainty of the EoS, except at the lower temperature of 260 K and pressures between 8 MPa and 13 MPa, where positive deviations of up to 0.19 % are observed. At pressures above 19 MPa, slight negative deviations exceeding the stated uncertainty of the EoS can also be seen at a few points at 300 K. Deviations from the GERG-2008 EoS (Figure 3(b)) exceed the claimed uncertainty of the EoS at 260 K for pressures between 4 MPa and 9 MPa (with positive deviations up to 0.20 %) and across all isotherms (except 350 K) for pressures above 12 MPa (260 K, 275 K, and 300 K) and above 17 MPa (325 K), (with negative deviations reaching -0.44 %). Finally, deviations from the improved GERG-2008 EoS (Figure 3(c)) fall within the stated uncertainty only for density data at 350 K. For other isotherms, negative deviations occur for pressures between 5 MPa and 20 MPa, reaching their highest magnitude of -0.38 % for 260 K at around 13 MPa. These results are significantly worse than the equivalent results of our previous work for the 10 % $H_2$-enriched NG derived from a NG with a significantly higher methane content (G 453) [32], where maximum relative deviations of approximately 0.20 % for any of the three EoS models were observed only at the lowest temperature (which was 250 K in that study).

Figure 4 displays the relative deviations of the experimental density data for the G 456 $H_2$-enriched NG mixture (G 432 + 20 % $H_2$) with respect to the three employed EoS models. Similar to the observations for the mixture G 455 (10 % $H_2$), deviations from the AGA8-DC92 EoS (Figure 4(a)) remain within the stated uncertainty, except at the lower temperature of 260 K and pressures between 7 MPa and 17 MPa, where positive deviations of up to 0.28 % are noted. Likewise, as seen for the G 455 mixture, deviations from the GERG-2008 EoS (Figure 4(b)) exceed the claimed uncertainty at 260 K, with positive deviations up to 0.29 % for pressures below 11 MPa and negative deviations reaching -0.34 % for pressures above 15 MPa. At 300 K, negative deviations up to -0.25 % are observed for pressures above 15 MPa. Finally, deviations from the improved GERG-2008 EoS (Figure 4(c)) fall within the stated uncertainty over the $p$, $T$ range investigated, except at 260 K (above 11 MPa) and at 300 K (above 8 MPa), where negative deviations reach a maximum of -0.30 %. These deviations are, once again, significantly larger than those from our previous study on 20% $H_2$-enriched NG derived from the methane-rich gas mixture (G 454) [32], where maximum relative deviations of approximately 0.19% for any of the three EoS models were observed only at the lowest temperature (250 K in that study).

Figures 2 to 4 show that some relative deviations at the lower pressure do not approach zero as expected for ideal gas behavior. This may be due to sorption phenomena, which can slightly alter the mixture's composition and, consequently, the density measurements—particularly in complex mixtures like those studied here. The effects of adsorption and desorption on density measurements of multicomponent gas mixtures were extensively examined by Richter and Kleinrahm [58]. Although our measurement procedure minimizes these effects by evacuating and refilling the cell multiple times, the single-sinker densimeter used here cannot quantify this influence. Nonetheless, we are certain that these deviations are well below the experimental uncertainty, which for technical reasons inherent to the measurement procedure is relatively high at lower pressures and densities, as can be seen from the 'Error bars' on density data in Figures 2 to 4.

The values from the statistical analysis in Table 6 show that the AGA8-DC92 EoS model performs well for the $H_2$-free NG mixture and slightly less accurately for the 10 % and 20 % $H_2$-enriched mixtures, as indicated by increasing *AARD* values (0.040 %, 0.048 %, and 0.065 %) and maximum relative deviations (0.15 %, 0.19 %, and 0.28 %) when hydrogen is added. Only a few density data points at the lower temperature (260 K) have deviations exceeding the model's stated uncertainty.

The GERG-2008 and improved GERG-2008 EoS models produce higher deviations for the hydrogen-free mixture (G 432) than the AGA8-DC92, with maximum relative deviations of 0.21 % and 0.28 % compared to 0.15 %. Interestingly, these models handle the 20 % $H_2$-enriched mixture slightly better than the 10 % $H_2$-enriched mixture, with *AARD* values of 0.094 % and 0.077 % versus 0.11 % and 0.12 %, and maximum relative deviations (*MarRD*) of 0.34 % and 0.30 % versus 0.44 % and 0.38 %. For the 20 % $H_2$-enriched mixture, both the GERG and improved GERG models yield results similar to those of the AGA8-DC92 EoS.

Table 6 also includes results for the equivalent mixtures derived from a methane-rich NG of our previous work [32] for comparison. Overall, all three EoS models perform better with those mixtures derived from a NG of higher methane content than with the (methane-depleted) mixtures in this study, which contain significant amounts of heavier hydrocarbons.

## 4.2. Isothermal compressibility derived from experimental density data

The experimental density values provide additional state properties through their derivatives by applying thermodynamic potentials. Figure 5 shows the partial derivatives of experimental density with respect to pressure, $\left.\frac{\partial \rho_{\text{exp}}}{\partial p}\right|_T$, while Table 7 lists the resultant isothermal compressibility values, $\kappa_T = \frac{1}{\rho_{\text{exp}}}\left.\frac{\partial \rho_{\text{exp}}}{\partial p}\right|_T$, for the three studied mixtures. These derivatives were calculated using cubic spline interpolation on the measured density data, with values at the maximum and minimum pressures of each isotherm excluded from the analysis. As temperature decreases, the partial derivatives exhibit a more convex shape, which flattens as hydrogen content increases across the studied pressure and temperature ranges. The $\kappa_T$ values span from (0.0361 to 0.5427) $\text{MPa}^{-1}$ at 250 K for the $H_2$-free NG mixture (G 432), decreasing as both temperature and hydrogen concentration rise. A similar qualitative behavior, with comparable $\kappa_T$ values, was observed in our previous work [32] for the NG mixture (mainly methane, > 97 %) and the two derived mixtures with 10 % and 20 % hydrogen.

The solid lines in Figure 5, representing the values of $\left.\frac{\partial \rho_{\text{EoS}}}{\partial p}\right|_T$ predicted by the improved GERG-2008 EoS, closely match those values derived from the experimental densities obtained in this work. Statistical analysis of the relative deviations shows *AARD* values of 0.23 %, 0.29 %, and 0.29 % and consequent *MaxRD* values of 1.9 %, 1.6 %, and 1.3 % for the $H_2$-free (G 432), 10 % $H_2$ (G 455), and 20 % $H_2$ (G 456) mixtures, respectively. All deviations are located within the estimated expanded ($k = 2$) uncertainty of $\kappa_T$, $U_r(\kappa_T) = 0.7$ %. The same was observed for the mixtures analyzed in our previous work [32], but in that case relative deviations were even smaller with *AARD* values of 0.18 %, 0.14 %, and 0.13 % and *MaxRD* values of 2.0 %, 1.7 %, and 1.3 % for the $H_2$-free (G 431), 10 % $H_2$ (G 453) and 20 % $H_2$ (G 454) mixtures, respectively. Comparison with the other EoS considered in this work gave similar results. Statistical analysis of the relative deviations against values predicted by AGA8-DC92 and GERG-2008 EoS shows AARD values below 0.21 % and 0.26 %, and consequent MaxRD values below 2.0 % and 1.9 %, respectively, for the three mixtures.

# 5. Conclusions

This study evaluates the accuracy of reference equations of state (EoS) commonly used for NG when applied to hydrogen-enriched NG (H2NG) mixtures. To achieve this, density measurements were taken for three synthetic NG-related mixtures across a temperature range of (260 to 350) K and pressures up to 20 MPa, using a high-precision single-sinker magnetic suspension densimeter. The first mixture is a high-calorific, 11-component NG blend with substantial ethane and propane content. The other two mixtures were created by adding hydrogen to the base mixture, resulting in compositions with a nominal hydrogen content of 10 % and 20 % (mol/mol), respectively. All mixtures were gravimetrically prepared to maximize composition accuracy.

The experimental density data was compared to densities predicted by three reference EoS: AGA8-DC92, GERG-2008, and an improved version of GERG-2008. For the $H_2$-free NG mixture (G 432), AGA8-DC92 generally remained within its stated uncertainty (0.1 %), except at the lowest temperature (260 K) and moderate pressures, while both GERG-2008 and its improved version displayed larger deviations that became more pronounced especially at lower temperatures and higher pressures. For the 10 % $H_2$-enriched mixture (G 455), AGA8-DC92 mostly remained within its uncertainty range, though it showed some deviation at low temperatures and high pressures. GERG-2008 and improved GERG-2008 exceeded their uncertainty limits (0.1 %) at lower temperatures and some pressure levels, with deviations reaching up to -0.44 %. In the 20 % $H_2$-enriched mixture (G 456), deviations from all three models exceeded the stated uncertainties at low temperatures and higher pressures, following a similar pattern to the 10% mixture.

AGA8-DC92 exhibited smaller deviations across both $H_2$-free and $H_2$-enriched mixtures than the GERG models. Maximum relative deviations increased for the AGA8-DC92 EoS as hydrogen content increased. However, the GERG models showed slightly improved performance for the 20% $H_2$-enriched mixture over the 10 % mixture. Compared with the findings from our previous study [32], we can say that, in general, EoS models performed best with "simpler", methane-rich mixtures. However, for mixtures that deviate from this composition, such as those with significant ethane and propane content studied in this work, EoS accuracy decreased, especially at lower temperatures and higher pressures.

Experimental density data for multicomponent H2NG mixtures are crucial for refining and validating reference EoS models, which are essential for accurate system design, operation, and custody transfer processes. This will become more relevant because a higher diversity in the composition of energy gases can be expected. Experimental data on binary mixtures of ethane or propane with hydrogen remain crucial for the refinement of these EoS.

**Acknowledgments**

This study was supported by the European Metrology Programme for Innovation and Research (EMPIR), Funder ID: 10.13039/100014132, Grant No. 19ENG03 MefHySto, as well as by the Regional Government of Castilla y León (Junta de Castilla y León), the Ministry of Science and Innovation MCIN, and the European Union NextGenerationEU/PRTR, under project C17.I01.P01.S21.

**Figures**

**Figure 1.**

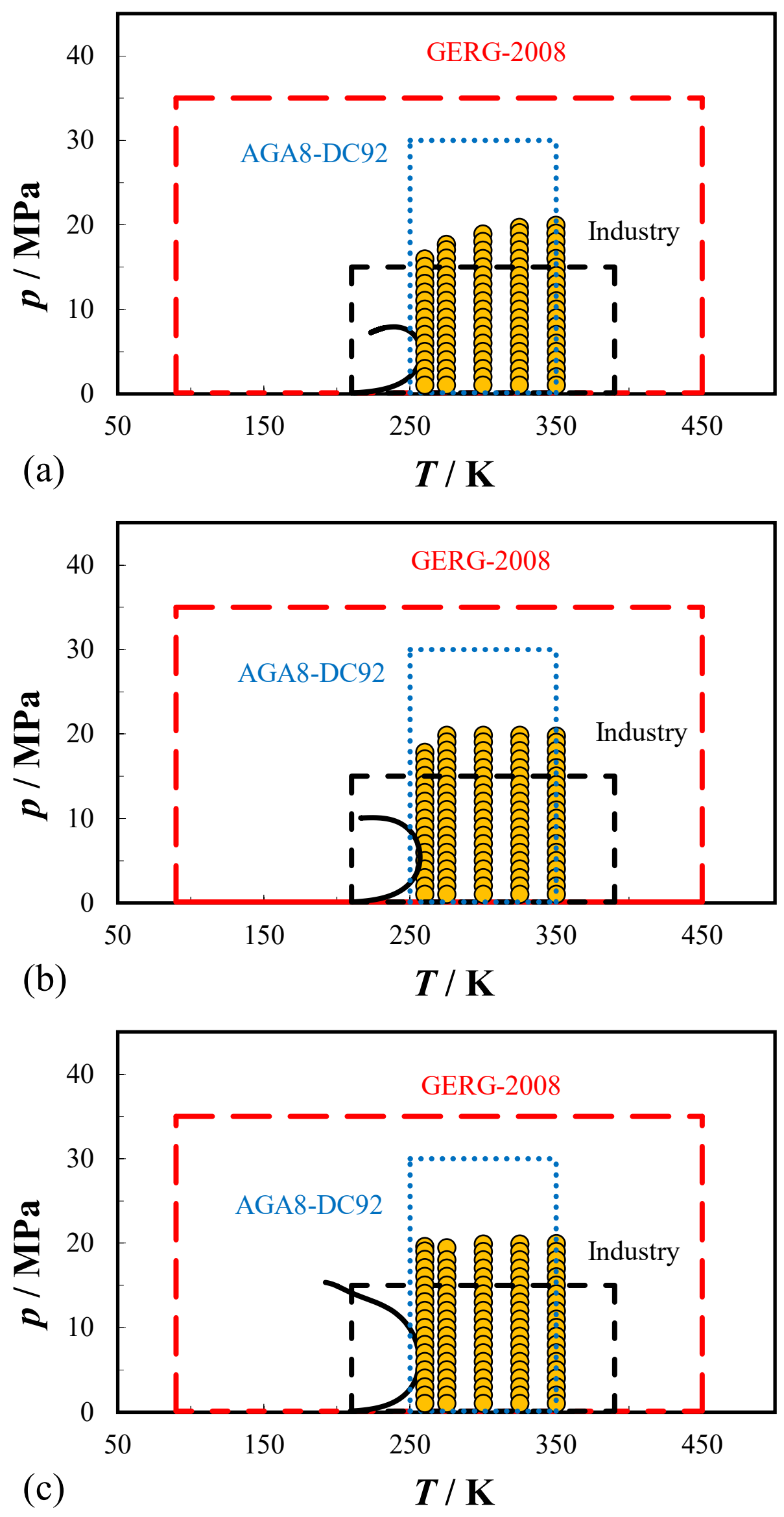

**Figure 1.** *p*, *T*-phase diagram with experimental points (●) and the calculated phase envelope (solid line) using the improved GERG EoS [52–54] for: a) NG mixture G 432 ($H_2$-free), b) H2NG mixture G 455 (10 % $H_2$), and c) H2NG gas mixture G 456 (20 % $H_2$). The marked temperature and pressure ranges indicate the validity of the AGA8-DC92 EoS [19] and GERG-2008 EoS [20,21], as well as the area of interest for the gas industry.

**Figure 2.**

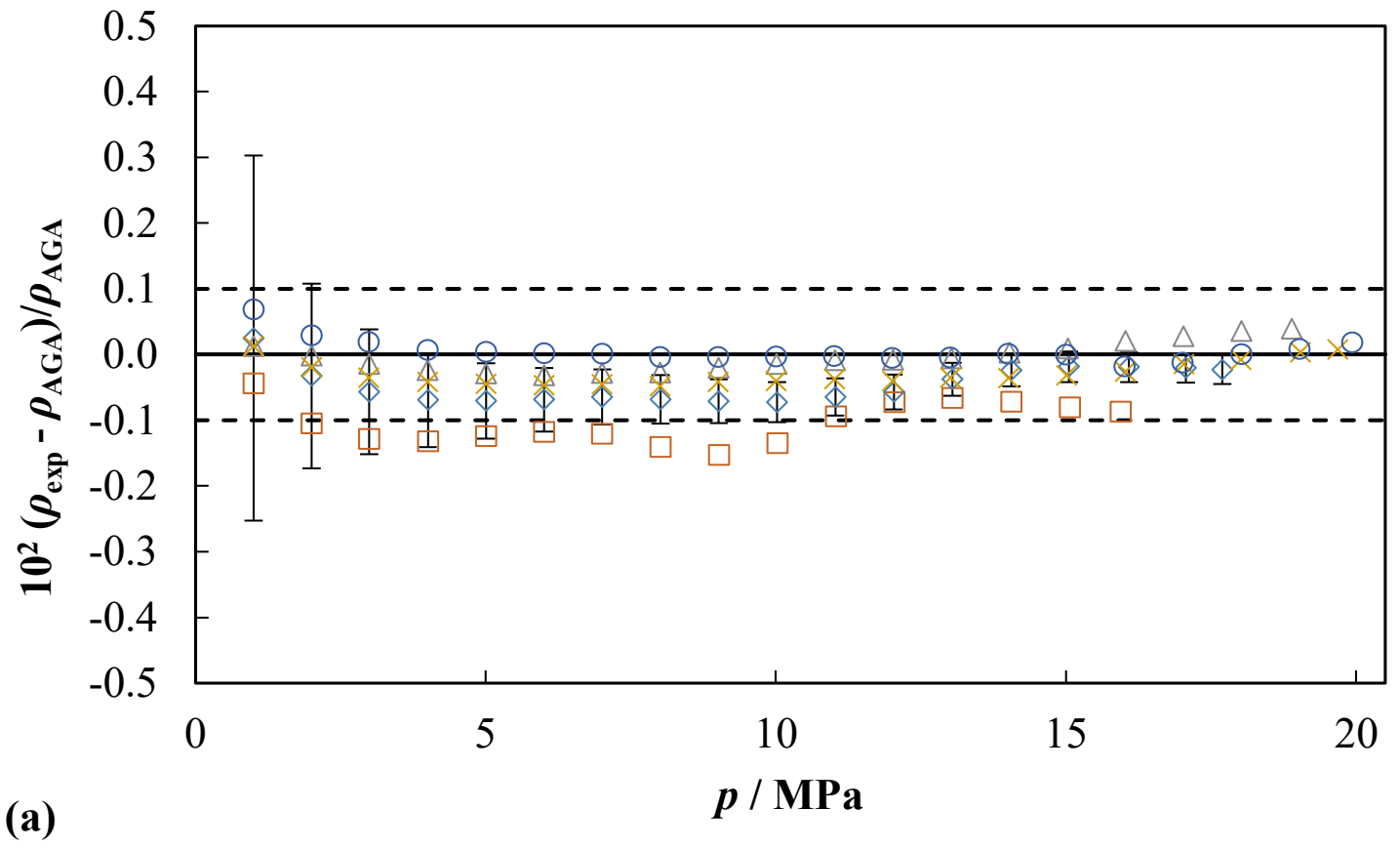

(a)

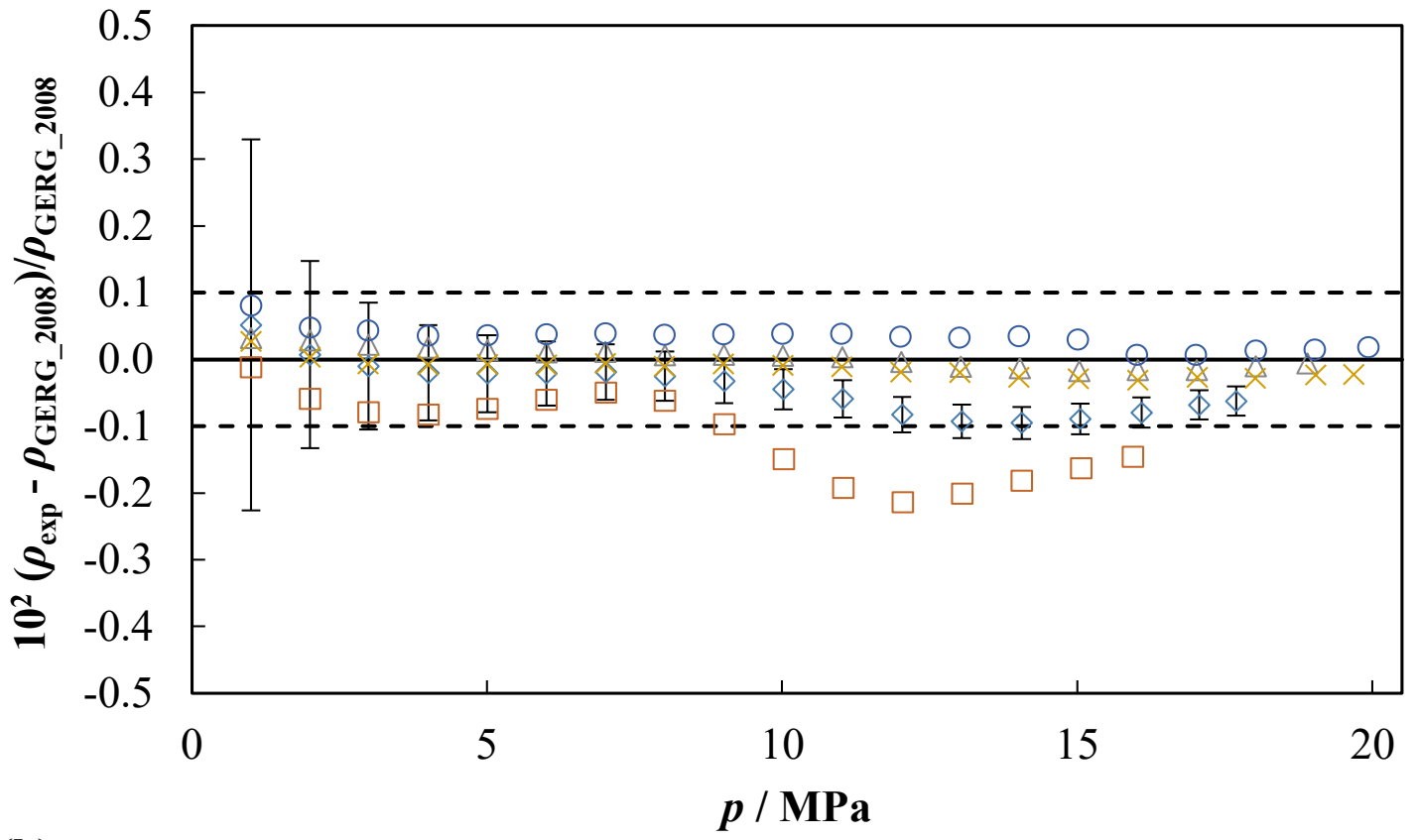

(b)

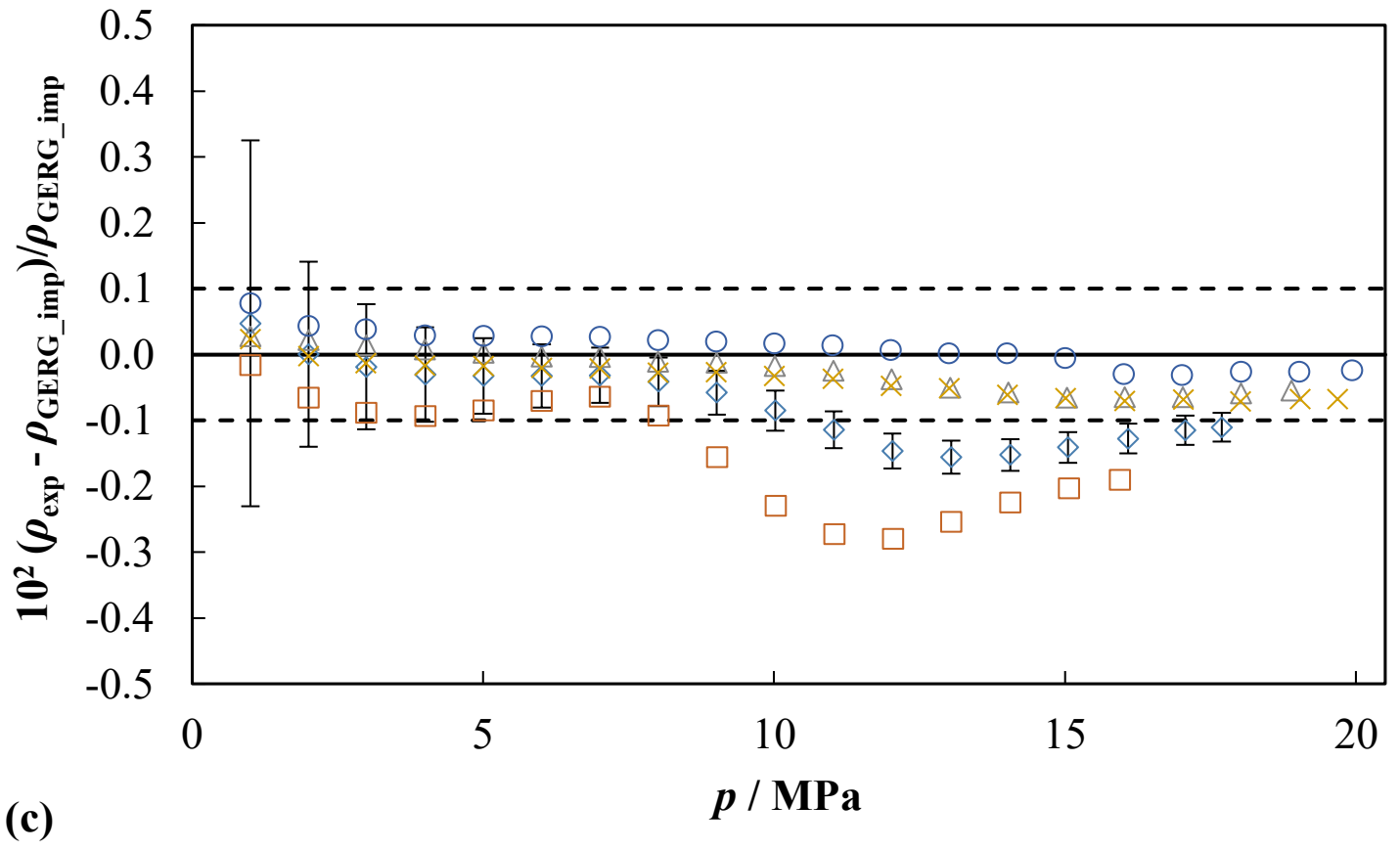

(c)

**Figure 2.** Relative deviations in density of experimental ($p$, $\rho_{exp}$, $T$) data of the ($H_2$-free) NG mixture G 432 from density values $\rho$ calculated from (a) AGA8-DC92 EoS [19], (b) GERG-2008 EoS [20,21], and (c) improved GERG-2008 EoS [52–54], as a function of the pressure for different temperatures: □ 260 K, ◇ 275 K, △ 300 K, × 325 K, ○ 350 K. Dashed lines represent the expanded ($k = 2$) uncertainty of the EoS. Error bars on the 275-K data set show the expanded ($k = 2$) uncertainty of the experimental density.

**Figure 3.**

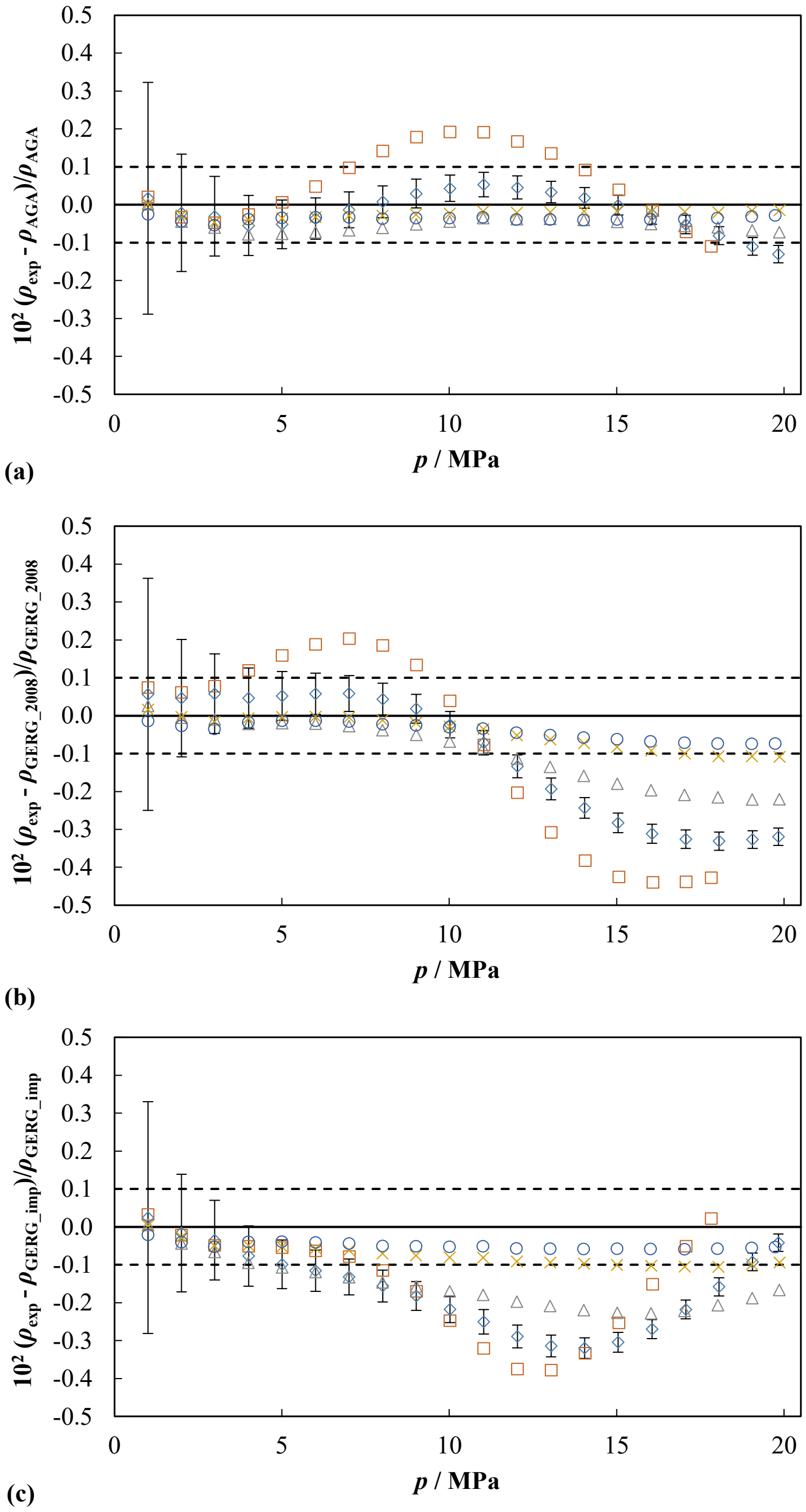

**Figure 3.** Relative deviations in density of experimental ($p$, $\rho_{exp}$, $T$) data of the H2NG mixture G 455 (10 % $H_2$) from density values $\rho$ calculated from (a) AGA8-DC92 EoS [19], (b) GERG-2008 EoS [20,21], and (c) improved GERG-2008 EoS [52–54], as a function of the pressure for different temperatures: □ 260 K, ◇ 275 K, △ 300 K, × 325 K, ○ 350 K. Dashed lines indicate the expanded ($k$ = 2) uncertainty of the EoS. Error bars on the 275-K data set show the expanded ($k$ = 2) uncertainty of the experimental density.

**Figure 4.**

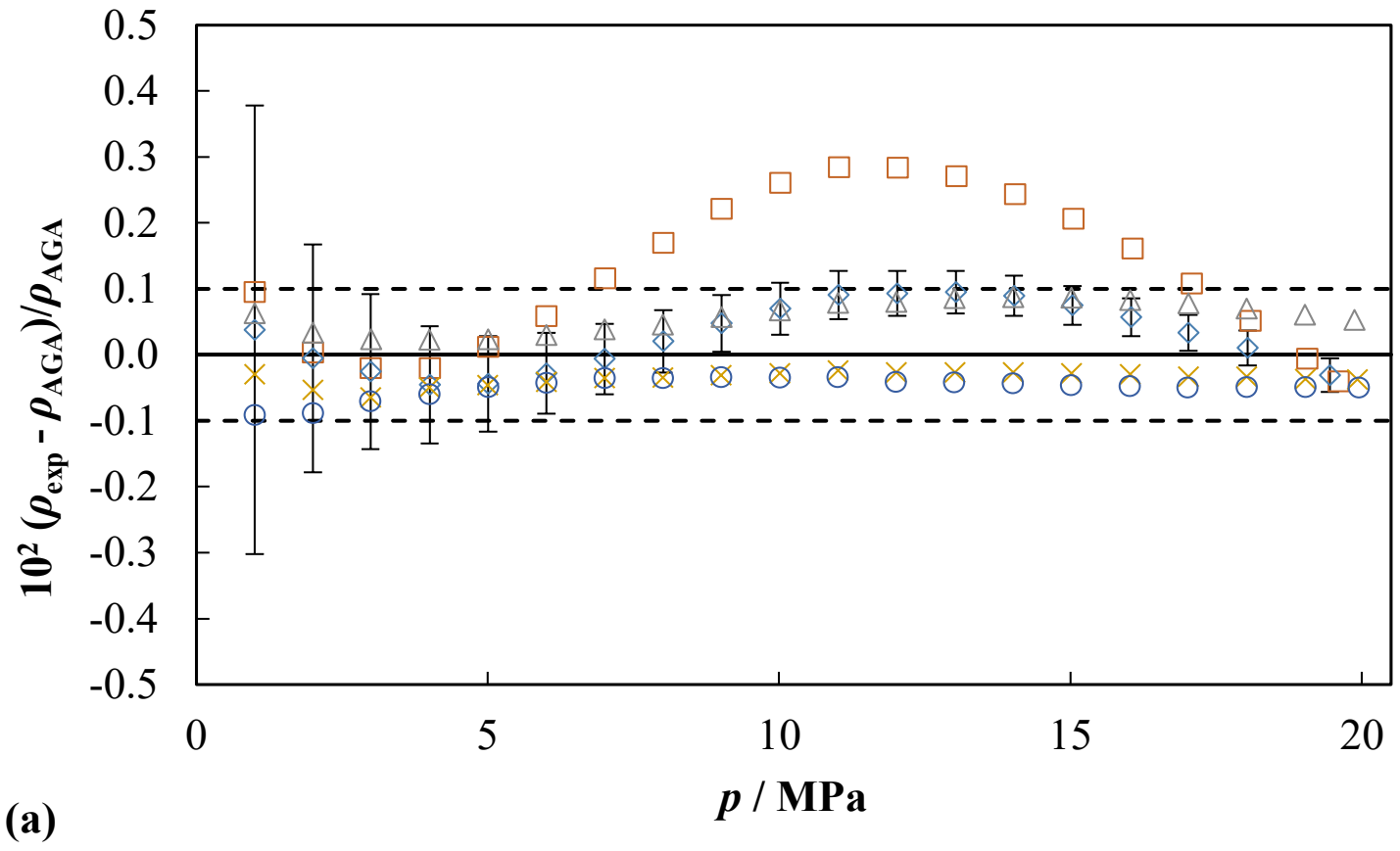

**(a)**

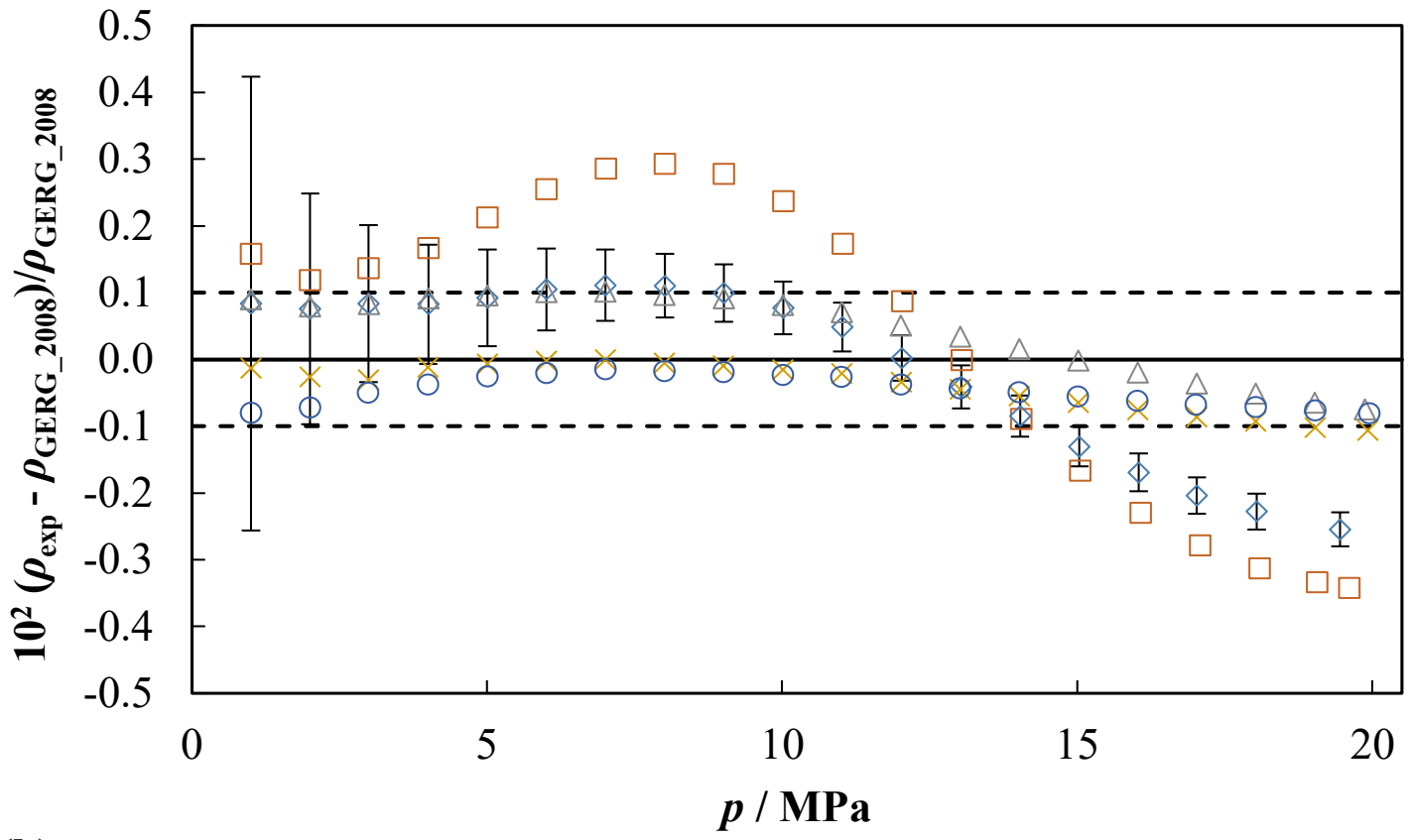

**(b)**

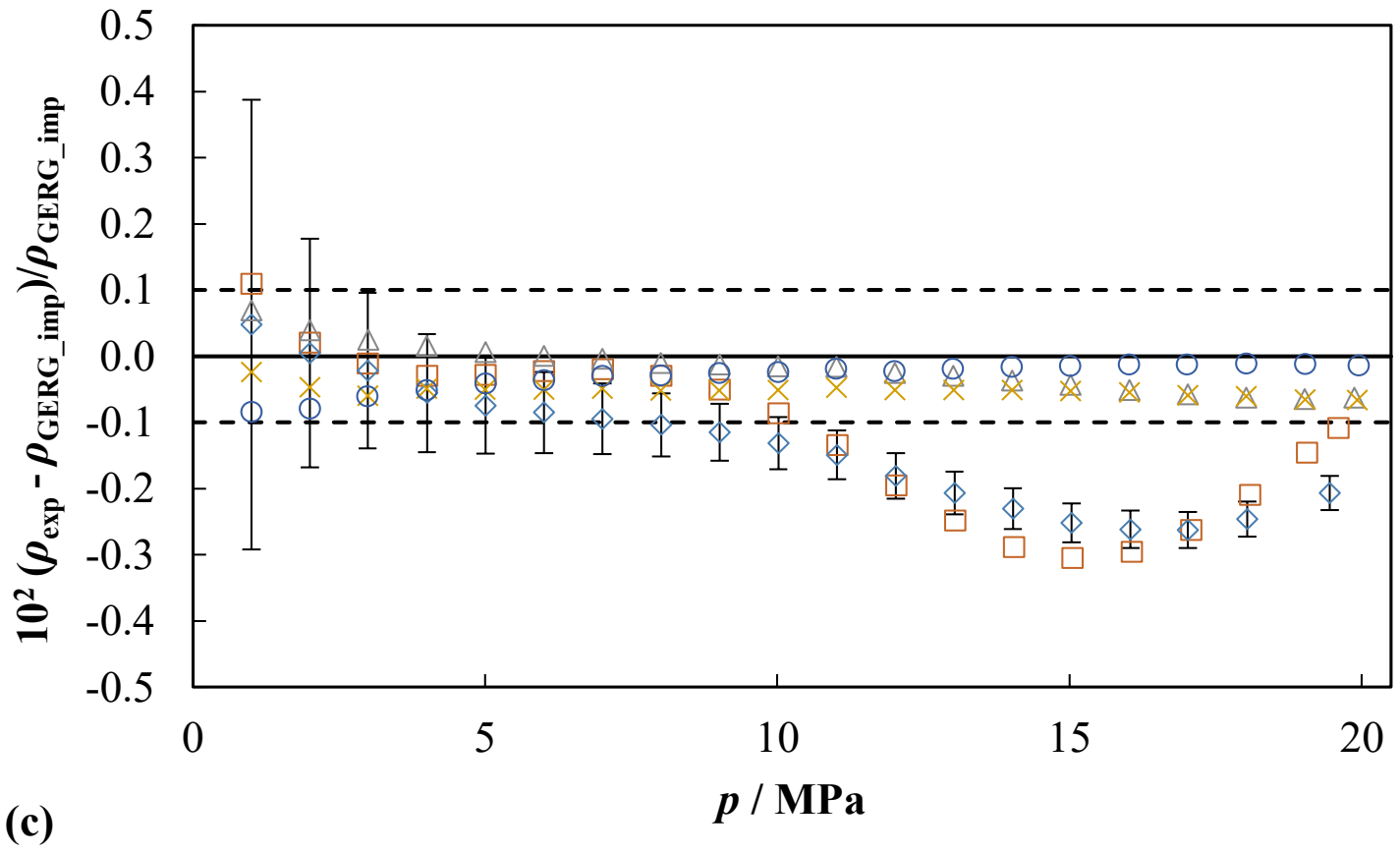

**(c)**

**Figure 4.** Relative deviations in density of experimental ($p$, $\rho_{exp}$, $T$) data of the H2NG mixture G 456 (20 % $H_2$) from density values $\rho$ calculated from (a) AGA8-DC92 EoS [19], (b) GERG-2008 EoS [20,21], and (c) improved GERG-2008 EoS [52–54], as a function of the pressure for different temperatures: □ 260 K, ◇ 275 K, △ 300 K, × 325 K, ○ 350 K. Dashed lines indicate the expanded ($k$ = 2) uncertainty of the EoS. Error bars on the 275-K data set show the expanded ($k$ = 2) uncertainty of the experimental density.

**Figure 5.**

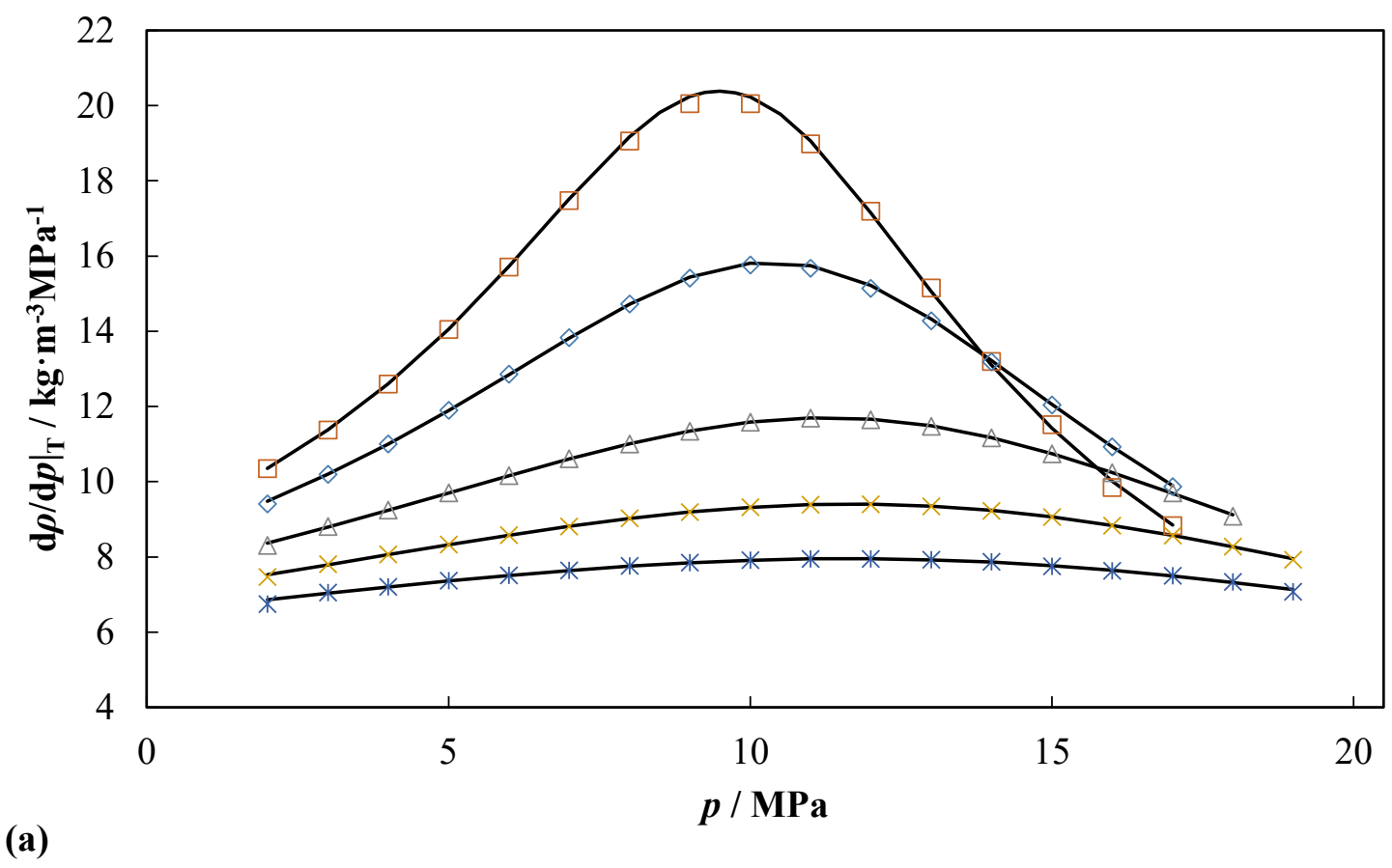

(a)

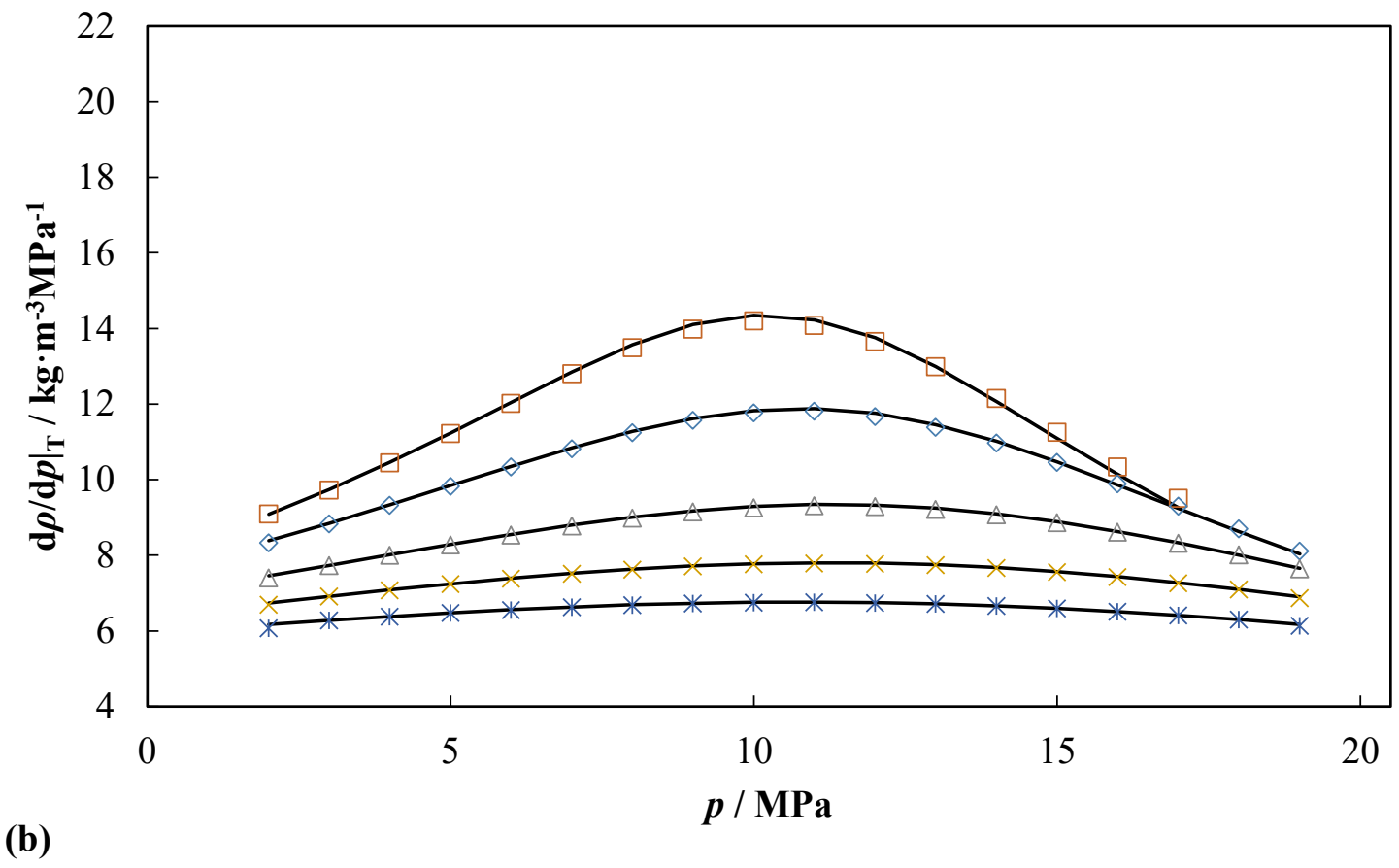

(b)

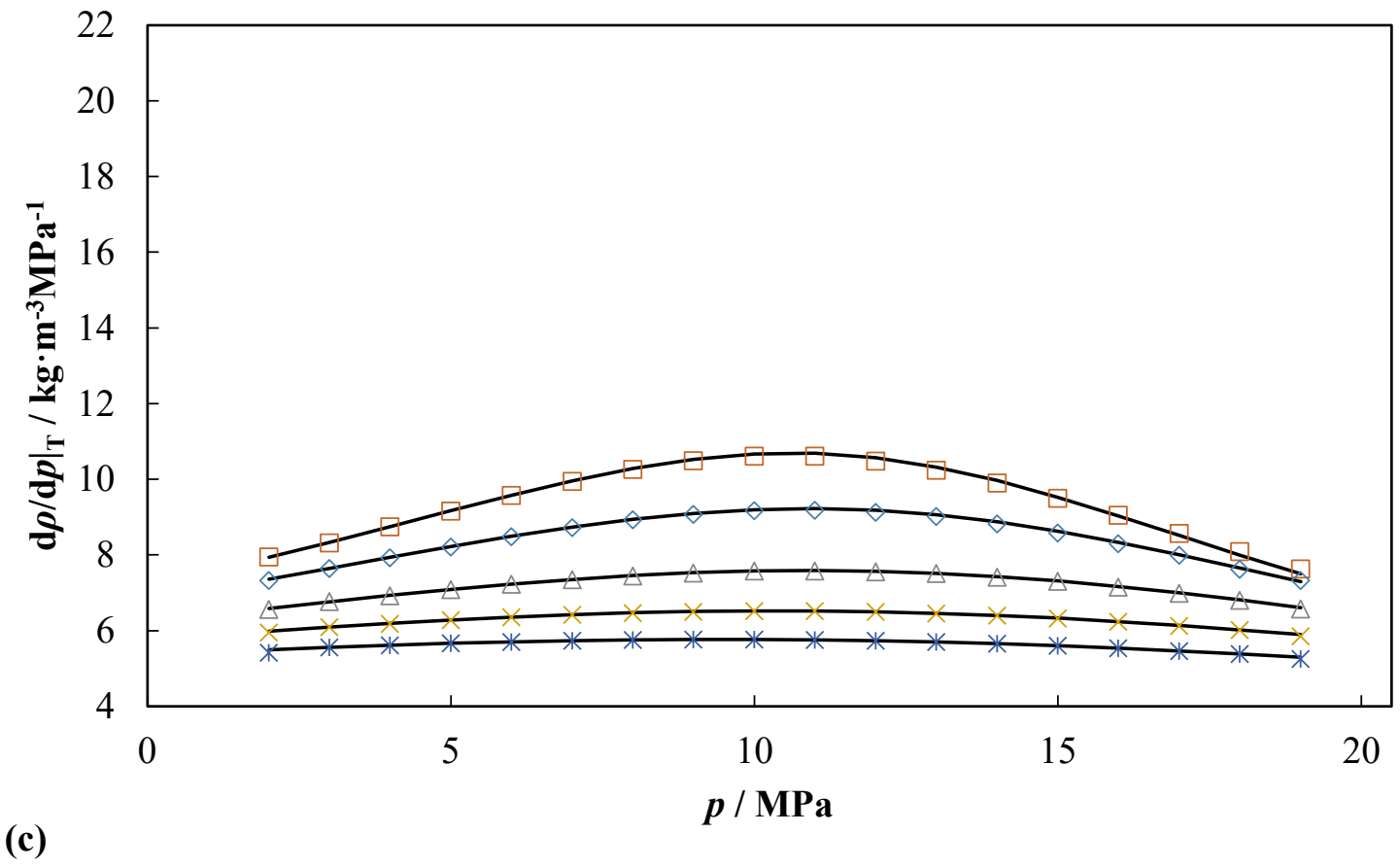

(c)

**Figure 5.** Derived $\left.\frac{\partial \rho_{\mathrm{exp}}}{\partial p}\right|_T$ values of (a) G 432, (b) G 455 (10 % $H_2$), and (c) G 456 (20 % $H_2$) mixtures as a function of the pressure for different temperatures: □ 260 K, ◇ 275 K, △ 300 K, × 325 K, ○ 350 K. Solid lines indicate the $\left.\frac{\partial \rho_{\mathrm{EoS}}}{\partial p}\right|_T$ values calculated from the improved GERG-2008 EoS [52–54].

**Tables**

**Table 1.** Composition of the reference NG mixtures studied in this work, with impurity compounds indicated in *italics*.

| Component | G 432 BAM no: 2031-200915 | | G 455 (G 432 + 10 % $H_2$) BAM no: 6079-210111 | | G 456 (G 432 + 20 % $H_2$) BAM no: 2059-210201 | |
|---|---|---|---|---|---|---|
| | $10^2\ x_i$ / mol/mol | $10^2\ U(x_i)$ / mol/mol | $10^2\ x_i$ / mol/mol | $10^2\ U(x_i)$ / mol/mol | $10^2\ x_i$ / mol/mol | $10^2\ U(x_i)$ / mol/mol |
| Methane | 85.0063 | 0.0026 | 76.5104 | 0.0031 | 68.0159 | 0.0027 |
| Hydrogen | – | – | 9.9864 | 0.0027 | 20.0039 | 0.0046 |
| Nitrogen | 0.95080 | 0.00018 | 0.85506 | 0.00016 | 0.76016 | 0.00015 |
| Carbon dioxide | 1.44823 | 0.00016 | 1.30441 | 0.00016 | 1.16003 | 0.00015 |
| Ethane | 8.99177 | 0.00072 | 8.11044 | 0.00073 | 7.18322 | 0.00072 |
| Propane | 3.00256 | 0.00051 | 2.69314 | 0.00051 | 2.39645 | 0.00050 |
| *n*-Butane | 0.19994 | 0.00012 | 0.18016 | 0.00010 | 0.160109 | 0.000091 |
| Isobutane | 0.200443 | 0.000074 | 0.180025 | 0.000059 | 0.160276 | 0.000053 |
| *n*-Pentane | 0.050054 | 0.000021 | 0.044993 | 0.000018 | 0.040018 | 0.000016 |
| Isopentane | 0.049929 | 0.000020 | 0.044914 | 0.000018 | 0.039882 | 0.000016 |
| Neopentane | 0.050035 | 0.000031 | 0.045022 | 0.000028 | 0.040009 | 0.000025 |
| *n*-Hexane | 0.049965 | 0.000016 | 0.044995 | 0.000014 | 0.039986 | 0.000013 |
| *Oxygen* | *0.0000120* | *0.0000070* | *0.0000130* | *0.0000061* | *0.0000150* | *0.0000078* |
| *Hydrogen* | *0.0000020* | *0.0000011* | - | - | - | - |
| *Carbon monoxide* | *0.00000100* | *0.00000078* | *0.0000020* | *0.0000014* | *0.0000030* | *0.0000026* |
| *Propene* | *0.0000030* | *0.0000035* | *0.0000030* | *0.0000035* | *0.0000020* | *0.0000023* |
| *Ethylene* | *0.0000010* | *0.0000012* | *0.0000010* | *0.0000012* | *0.0000010* | *0.0000012* |
| *Nitric oxide* | *0.00000010* | *0.00000012* | *0.00000010* | *0.00000012* | *0.00000010* | *0.00000012* |
| Normalized composition without impurities | | | | | | |
| Methane | 85.0063 | 0.0026 | 76.5104 | 0.0031 | 68.0159 | 0.0027 |
| Hydrogen | – | – | 9.9864 | 0.0027 | 20.0039 | 0.0046 |
| Nitrogen | 0.95080 | 0.00018 | 0.85506 | 0.00016 | 7.18322 | 0.00072 |
| Carbon dioxide | 1.44823 | 0.00016 | 1.30441 | 0.00016 | 2.39645 | 0.00050 |
| Ethane | 8.99177 | 0.00072 | 8.11044 | 0.00073 | 1.16003 | 0.00015 |
| Propane | 3.00256 | 0.00051 | 2.69314 | 0.00051 | 0.76016 | 0.00015 |
| *n*-Butane | 0.19994 | 0.00012 | 0.18016 | 0.00010 | 0.160276 | 0.000053 |
| Isobutane | 0.200443 | 0.000074 | 0.180025 | 0.000059 | 0.160109 | 0.000091 |
| *n*-Pentane | 0.050054 | 0.000021 | 0.044993 | 0.000018 | 0.040018 | 0.000016 |
| Isopentane | 0.049929 | 0.000020 | 0.044914 | 0.000018 | 0.040009 | 0.000025 |
| *n*-Hexane | 0.049965 | 0.000016 | 0.044995 | 0.000014 | 0.039882 | 0.000016 |
| Neopentane | 0.050035 | 0.000031 | 0.045022 | 0.000028 | 0.039986 | 0.000013 |

| | | | |
|---|---|---|---|
| Molar mass $M$ g / mol | 18.954 | 17.260 | 15.566 |
| Normalized density $\rho_n$ kg / m$^3$ | 0.80379 | 0.73154 | 0.65940 |
| Higher Heating Value $HHV$ MJ / m$^3$ | 41.726 | 38.744 | 35.768 |
| Wobbe index $W_s$ MJ / m$^3$ | 51.523 | 50.148 | 48.763 |

**Table 2.** Contributions to the overall expanded ($k = 2$) uncertainty in density, $U_T(\rho_{exp})$, for the three reference NG mixtures examined in this study.

| Source | Contribution ($k = 2$) | Units | Estimation in density ($k = 2$) | |
|---|---|---|---|---|
| | | | kg·m$^{-3}$ | % |
| | | G 432 | | |
| Temperature, $T$ | 0.015 | K | < 0.0028 | < 0.0019 |
| Pressure, $p$ | < 0.005 | MPa | (0.023–0.086) | (0.020–0.37) |
| Composition, $x_i$ | < 0.0004 | mol·mol$^{-1}$ | < 0.011 | < 0.0065 |
| Density, $\rho$ | (0.024–0.050) | kg·m$^{-3}$ | (0.024–0.050) | (0.021–0.036) |
| Sum | | | (0.033–0.096) | (0.030–0.50) |
| | | G 455 (G 432 + 10 % $H_2$) | | |
| Temperature, $T$ | 0.015 | K | < 0.0022 | < 0.0018 |
| Pressure, $p$ | < 0.005 | MPa | (0.022–0.062) | (0.020–0.37) |
| Composition, $x_i$ | < 0.0004 | mol·mol$^{-1}$ | < 0.015 | < 0.0088 |
| Density, $\rho$ | (0.024–0.046) | kg·m$^{-3}$ | (0.024–0.046) | (0.023–0.039) |
| Sum | | | (0.032–0.073) | (0.031–0.53) |
| | | G 456 (G 432 + 20 % $H_2$) | | |
| Temperature, $T$ | 0.015 | K | < 0.0018 | < 0.0016 |
| Pressure, $p$ | < 0.005 | MPa | (0.019–0.047) | (0.020–0.37) |
| Composition, $x_i$ | < 0.0004 | mol·mol$^{-1}$ | < 0.017 | < 0.010 |
| Density, $\rho$ | (0.023–0.043) | kg·m$^{-3}$ | (0.023–0.043) | (0.024–0.44) |
| Sum | | | (0.031–0.061) | (0.033–0.57) |

**Table 3.** Experimental ($p$, $\rho_{exp}$, $T$) measurements for the ($H_2$-free) NG mixture G 432, absolute and relative expanded ($k$ = 2) uncertainty in density, $U(\rho_{exp})$, relative deviations from the density given by the AGA8-DC92 EoS [19], $\rho_{AGA8\text{-}DC92}$, the GERG-2008 EoS [20,21], $\rho_{GERG\text{-}2008}$, and the improved GERG-2008 EoS [52–54], $\rho_{GERG\text{-}improved}$.

| $T$ / K[(a)] | $p$ / MPa[(a)] | $\rho_{exp}$ / kg·m$^{-3}$[(a)] | $U(\rho_{exp})$ / kg·m$^{-3}$ | $10^2\ U(\rho_{exp})/\rho_{exp}$ | $10^2\ (\rho_{exp} - \rho_{AGA8\text{-}DC92})/\rho_{AGA8\text{-}DC92}$ | $10^2\ (\rho_{exp} - \rho_{GERG\text{-}2008})/\rho_{GERG\text{-}2008}$ | $10^2\ (\rho_{exp} - \rho_{GERG\text{-}improved})/\rho_{GERG\text{-}improved}$ |
|---|---|---|---|---|---|---|---|
| | | | | 260.000 K | | | |
| 260.172 | 15.94387 | 234.701 | 0.050 | 0.021 | −0.087 | −0.146 | −0.190 |
| 260.173 | 15.06566 | 225.286 | 0.049 | 0.022 | −0.081 | −0.163 | −0.203 |
| 260.173 | 14.06213 | 213.074 | 0.047 | 0.022 | −0.072 | −0.182 | −0.224 |
| 260.172 | 13.04828 | 198.893 | 0.046 | 0.023 | −0.067 | −0.201 | −0.254 |
| 260.171 | 12.04212 | 182.780 | 0.044 | 0.024 | −0.073 | −0.214 | −0.280 |
| 260.174 | 11.03616 | 164.678 | 0.042 | 0.025 | −0.094 | −0.193 | −0.272 |
| 260.171 | 10.03085 | 145.054 | 0.039 | 0.027 | −0.135 | −0.149 | −0.230 |
| 260.171 | 9.02392 | 124.786 | 0.037 | 0.030 | −0.153 | −0.097 | −0.155 |
| 260.173 | 8.01756 | 105.026 | 0.035 | 0.033 | −0.141 | −0.062 | −0.093 |
| 260.173 | 7.01074 | 86.610 | 0.033 | 0.038 | −0.120 | −0.050 | −0.064 |
| 260.173 | 6.00939 | 70.010 | 0.031 | 0.044 | −0.118 | −0.060 | −0.070 |
| 260.175 | 5.00703 | 55.129 | 0.029 | 0.053 | −0.125 | −0.074 | −0.085 |
| 260.174 | 4.00615 | 41.834 | 0.028 | 0.066 | −0.132 | −0.083 | −0.093 |
| 260.176 | 2.99405 | 29.750 | 0.026 | 0.088 | −0.129 | −0.079 | −0.088 |
| 260.170 | 2.00451 | 19.044 | 0.025 | 0.132 | −0.105 | −0.059 | −0.065 |
| 260.180 | 1.00292 | 9.138 | 0.024 | 0.262 | −0.044 | −0.012 | −0.016 |
| | | | | 275.000 K | | | |
| 275.061 | 17.69592 | 221.059 | 0.048 | 0.022 | −0.023 | −0.063 | −0.110 |
| 275.061 | 17.06480 | 215.039 | 0.047 | 0.022 | −0.020 | −0.068 | −0.115 |
| 275.060 | 16.08785 | 204.939 | 0.046 | 0.023 | −0.019 | −0.080 | −0.127 |
| 275.060 | 15.05121 | 193.104 | 0.045 | 0.023 | −0.019 | −0.089 | −0.141 |
| 275.060 | 14.05879 | 180.626 | 0.043 | 0.024 | −0.024 | −0.095 | −0.152 |
| 275.061 | 13.04325 | 166.705 | 0.042 | 0.025 | −0.038 | −0.093 | −0.155 |
| 275.059 | 12.03336 | 151.847 | 0.040 | 0.026 | −0.057 | −0.083 | −0.146 |
| 275.060 | 11.02685 | 136.321 | 0.038 | 0.028 | −0.065 | −0.059 | −0.114 |
| 275.059 | 10.02282 | 120.507 | 0.037 | 0.030 | −0.073 | −0.045 | −0.085 |
| 275.059 | 9.01787 | 104.818 | 0.035 | 0.033 | −0.071 | −0.032 | −0.058 |
| 275.058 | 8.01330 | 89.673 | 0.033 | 0.037 | −0.068 | −0.025 | −0.041 |
| 275.059 | 7.01084 | 75.371 | 0.031 | 0.042 | −0.065 | −0.019 | −0.031 |
| 275.059 | 6.00932 | 62.022 | 0.030 | 0.048 | −0.069 | −0.021 | −0.032 |
| 275.060 | 5.00667 | 49.629 | 0.029 | 0.058 | −0.070 | −0.021 | −0.032 |
| 275.059 | 4.00525 | 38.178 | 0.027 | 0.071 | −0.069 | −0.020 | −0.030 |
| 275.058 | 2.99255 | 27.463 | 0.026 | 0.095 | −0.057 | −0.010 | −0.018 |
| 275.058 | 2.00274 | 17.741 | 0.025 | 0.140 | −0.033 | 0.007 | 0.001 |
| 275.060 | 1.00263 | 8.587 | 0.024 | 0.278 | 0.025 | 0.052 | 0.047 |
| | | | | 300.000 K | | | |
| 300.052 | 18.89613 | 191.689 | 0.045 | 0.023 | 0.039 | −0.006 | −0.054 |
| 300.053 | 18.02313 | 183.947 | 0.044 | 0.024 | 0.035 | −0.010 | −0.059 |
| 300.050 | 17.02221 | 174.545 | 0.043 | 0.025 | 0.027 | −0.016 | −0.065 |
| 300.052 | 16.02865 | 164.658 | 0.042 | 0.025 | 0.021 | −0.016 | −0.065 |
| 300.048 | 15.03453 | 154.249 | 0.040 | 0.026 | 0.009 | −0.018 | −0.065 |
| 300.050 | 14.02944 | 143.258 | 0.039 | 0.027 | 0.003 | −0.013 | −0.057 |
| 300.048 | 13.02806 | 131.933 | 0.038 | 0.029 | −0.006 | −0.011 | −0.050 |
| 300.052 | 12.02303 | 120.313 | 0.037 | 0.030 | −0.008 | −0.004 | −0.037 |
| 300.046 | 11.01692 | 108.579 | 0.035 | 0.032 | −0.009 | 0.003 | −0.024 |
| 300.048 | 10.01484 | 96.916 | 0.034 | 0.035 | −0.015 | 0.005 | −0.017 |
| 300.047 | 9.01055 | 85.404 | 0.033 | 0.038 | −0.021 | 0.007 | −0.012 |
| 300.042 | 8.01027 | 74.223 | 0.031 | 0.042 | −0.028 | 0.006 | −0.011 |
| 300.042 | 7.00785 | 63.392 | 0.030 | 0.048 | −0.029 | 0.011 | −0.004 |
| 300.041 | 6.00685 | 52.999 | 0.029 | 0.055 | −0.032 | 0.010 | −0.003 |
| 300.048 | 5.00506 | 43.053 | 0.028 | 0.065 | −0.029 | 0.014 | 0.003 |
| 300.050 | 4.00406 | 33.577 | 0.027 | 0.080 | −0.024 | 0.018 | 0.008 |
| 300.061 | 2.99033 | 24.441 | 0.026 | 0.105 | −0.015 | 0.022 | 0.015 |
| 300.059 | 2.00230 | 15.968 | 0.025 | 0.155 | −0.001 | 0.028 | 0.022 |
| 299.994 | 1.00209 | 7.801 | 0.024 | 0.305 | 0.013 | 0.032 | 0.028 |
| | | | | 325.000 K | | | |
| 325.008 | 19.68452 | 168.816 | 0.042 | 0.025 | 0.008 | −0.023 | −0.067 |
| 325.013 | 19.03937 | 163.762 | 0.042 | 0.025 | 0.004 | −0.023 | −0.067 |
| 325.013 | 18.01498 | 155.457 | 0.041 | 0.026 | −0.008 | −0.029 | −0.071 |
| 325.020 | 17.03292 | 147.195 | 0.040 | 0.027 | −0.015 | −0.027 | −0.068 |
| 325.016 | 16.02401 | 138.423 | 0.039 | 0.028 | −0.026 | −0.031 | −0.070 |
| 325.018 | 15.01422 | 129.394 | 0.038 | 0.029 | −0.032 | −0.029 | −0.066 |
| 325.020 | 14.01120 | 120.224 | 0.037 | 0.030 | −0.036 | −0.027 | −0.061 |
| 325.028 | 13.01072 | 110.933 | 0.036 | 0.032 | −0.035 | −0.020 | −0.051 |
| 325.025 | 12.00972 | 101.554 | 0.034 | 0.034 | −0.041 | −0.019 | −0.048 |
| 325.025 | 11.01092 | 92.178 | 0.033 | 0.036 | −0.038 | −0.011 | −0.037 |
| 325.028 | 10.00864 | 82.804 | 0.032 | 0.039 | −0.040 | −0.009 | −0.032 |
| 325.031 | 9.00811 | 73.544 | 0.031 | 0.043 | −0.042 | −0.006 | −0.027 |
| 325.022 | 8.00796 | 64.437 | 0.030 | 0.047 | −0.048 | −0.010 | −0.028 |
| 325.022 | 7.00712 | 55.515 | 0.029 | 0.053 | −0.046 | −0.006 | −0.021 |
| 325.021 | 6.00600 | 46.809 | 0.028 | 0.060 | −0.047 | −0.008 | −0.020 |
| 325.020 | 5.00509 | 38.351 | 0.027 | 0.071 | −0.045 | −0.008 | −0.017 |
| 325.017 | 4.00388 | 30.149 | 0.026 | 0.087 | −0.042 | −0.008 | −0.016 |
| 325.023 | 2.98813 | 22.100 | 0.025 | 0.115 | −0.036 | −0.007 | −0.013 |
| 325.033 | 2.00275 | 14.555 | 0.025 | 0.169 | −0.020 | 0.003 | −0.002 |
| 325.039 | 1.00275 | 7.160 | 0.024 | 0.331 | 0.012 | 0.027 | 0.023 |
| | | | | 350.000 K | | | |
| 350.025 | 19.92958 | 149.082 | 0.040 | 0.027 | 0.018 | 0.018 | −0.024 |
| 350.028 | 19.02344 | 142.700 | 0.039 | 0.027 | 0.008 | 0.015 | −0.026 |
| 350.036 | 18.02717 | 135.497 | 0.038 | 0.028 | $< 0.001$ | 0.014 | −0.026 |
| 350.030 | 17.00874 | 127.951 | 0.037 | 0.029 | −0.013 | 0.007 | −0.031 |
| 350.034 | 16.00839 | 120.379 | 0.037 | 0.030 | −0.018 | 0.007 | −0.030 |
| 350.081 | 15.00598 | 112.656 | 0.036 | 0.032 | $< 0.001$ | 0.030 | −0.005 |
| 350.091 | 14.00636 | 104.848 | 0.035 | 0.033 | 0.001 | 0.035 | 0.002 |
| 350.086 | 13.00643 | 96.959 | 0.034 | 0.035 | −0.005 | 0.033 | 0.002 |
| 350.089 | 12.00505 | 89.014 | 0.033 | 0.037 | −0.005 | 0.034 | 0.007 |
| 350.088 | 11.00552 | 81.075 | 0.032 | 0.040 | −0.002 | 0.039 | 0.014 |
| 350.091 | 10.00554 | 73.146 | 0.031 | 0.043 | −0.003 | 0.038 | 0.017 |
| 350.091 | 9.00400 | 65.256 | 0.030 | 0.046 | −0.004 | 0.038 | 0.020 |
| 350.091 | 8.00542 | 57.467 | 0.029 | 0.051 | −0.004 | 0.037 | 0.022 |
| 350.092 | 7.00527 | 49.769 | 0.029 | 0.057 | 0.001 | 0.039 | 0.027 |
| 350.094 | 6.00494 | 42.191 | 0.028 | 0.066 | 0.002 | 0.038 | 0.028 |
| 350.093 | 5.00447 | 34.753 | 0.027 | 0.077 | 0.004 | 0.036 | 0.029 |
| 350.093 | 4.00306 | 27.461 | 0.026 | 0.095 | 0.007 | 0.035 | 0.029 |
| 350.095 | 2.98569 | 20.223 | 0.025 | 0.125 | 0.019 | 0.043 | 0.038 |
| 350.095 | 2.00252 | 13.394 | 0.024 | 0.182 | 0.029 | 0.048 | 0.044 |
| 350.095 | 1.00263 | 6.621 | 0.024 | 0.357 | 0.069 | 0.081 | 0.078 |

[a] Expanded uncertainties ($k$ = 2): $U(p > 3)/\text{MPa} = 7.5 \cdot 10^{-5} \cdot \frac{p}{\text{MPa}} + 4 \cdot 10^{-3}$; $U(p < 3)/\text{MPa} = 6.0 \cdot 10^{-5} \cdot \frac{p}{\text{MPa}} + 2 \cdot 10^{-3}$; $U(T)$ = 15 mK; $\frac{U(\rho)}{\text{kg·m}^{-3}} = 2.5 \cdot 10^{4} \frac{\chi_S}{m^3 kg^{-1}} + 1.1 \cdot 10^{-4} \cdot \frac{\rho}{\text{kg·m}^{-3}} + 2.3 \cdot 10^{-2}$.

**Table 4.** Experimental ($p$, $\rho_{exp}$, $T$) measurements for the ($H_2$-enriched) NG mixture G 455 (10 % $H_2$), absolute and relative expanded ($k$ = 2) uncertainty in density, $U(\rho_{exp})$, relative deviations from the density given by the AGA8-DC92 EoS [19], $\rho_{AGA8\text{-}DC92}$, the GERG-2008 EoS [20,21], $\rho_{GERG\text{-}2008}$, and the improved GERG-2008 EoS [52–54], $\rho_{GERG\text{-}improved}$.

| $T$ / K[(a)] | $p$ / MPa[(a)] | $\rho_{exp}$ / kg·m$^{-3}$[(a)] | $U(\rho_{exp})$ / kg·m$^{-3}$ | $10^2$ $U(\rho_{exp})/\rho_{exp}$ | $10^2$ ($\rho_{exp}$ − $\rho_{AGA8\text{-}DC92}$)/$\rho_{AGA8\text{-}DC92}$ | $10^2$ ($\rho_{exp}$ − $\rho_{GERG\text{-}2008}$)/ $\rho_{GERG\text{-}2008}$ | $10^2$ ($\rho_{exp}$ − $\rho_{GERG\text{-}improved}$)/ $\rho_{GERG\text{-}improved}$ |
|---|---|---|---|---|---|---|---|
| | | | | 260.000 K | | | |
| 260.175 | 17.82969 | 206.100 | 0.046 | 0.023 | −0.110 | −0.427 | 0.022 |
| 260.176 | 17.07126 | 199.200 | 0.046 | 0.023 | −0.071 | −0.438 | −0.051 |
| 260.178 | 16.07939 | 189.459 | 0.044 | 0.023 | −0.016 | −0.440 | −0.151 |
| 260.176 | 15.06483 | 178.600 | 0.043 | 0.024 | 0.040 | −0.425 | −0.253 |
| 260.177 | 14.05752 | 166.893 | 0.042 | 0.025 | 0.092 | −0.383 | −0.333 |
| 260.177 | 13.04719 | 154.256 | 0.040 | 0.026 | 0.136 | −0.308 | −0.378 |
| 260.177 | 12.03347 | 140.786 | 0.039 | 0.028 | 0.167 | −0.203 | −0.375 |
| 260.177 | 11.03276 | 126.920 | 0.037 | 0.029 | 0.192 | −0.077 | −0.321 |
| 260.178 | 10.02125 | 112.602 | 0.036 | 0.032 | 0.192 | 0.039 | −0.247 |
| 260.182 | 9.01879 | 98.457 | 0.034 | 0.035 | 0.179 | 0.134 | −0.170 |
| 260.179 | 8.01531 | 84.655 | 0.033 | 0.038 | 0.142 | 0.186 | −0.114 |
| 260.180 | 7.01157 | 71.449 | 0.031 | 0.043 | 0.098 | 0.203 | −0.078 |
| 260.179 | 6.00996 | 59.015 | 0.030 | 0.050 | 0.048 | 0.189 | −0.063 |
| 260.178 | 5.00795 | 47.378 | 0.028 | 0.060 | 0.007 | 0.159 | −0.055 |
| 260.179 | 4.00498 | 36.523 | 0.027 | 0.074 | −0.026 | 0.120 | −0.051 |
| 260.178 | 2.99199 | 26.321 | 0.026 | 0.098 | −0.047 | 0.078 | −0.048 |
| 260.180 | 2.00384 | 17.046 | 0.025 | 0.146 | −0.033 | 0.061 | −0.022 |
| 260.181 | 1.00333 | 8.264 | 0.024 | 0.288 | 0.021 | 0.074 | 0.033 |
| | | | | 275.000 K | | | |
| 275.149 | 19.83024 | 198.637 | 0.046 | 0.023 | −0.131 | −0.319 | −0.042 |
| 275.149 | 19.05955 | 192.537 | 0.045 | 0.023 | −0.110 | −0.327 | −0.092 |
| 275.149 | 18.05989 | 184.137 | 0.044 | 0.024 | −0.082 | −0.331 | −0.158 |
| 275.150 | 17.06344 | 175.188 | 0.043 | 0.024 | −0.052 | −0.326 | −0.218 |
| 275.150 | 16.05558 | 165.537 | 0.042 | 0.025 | −0.026 | −0.311 | −0.270 |
| 275.152 | 15.04279 | 155.246 | 0.041 | 0.026 | −0.001 | −0.283 | −0.304 |
| 275.151 | 14.04366 | 144.550 | 0.039 | 0.027 | 0.018 | −0.243 | −0.320 |
| 275.152 | 13.04090 | 133.347 | 0.038 | 0.029 | 0.034 | −0.193 | −0.314 |
| 275.153 | 12.03273 | 121.726 | 0.037 | 0.030 | 0.046 | −0.134 | −0.289 |
| 275.153 | 11.02767 | 109.934 | 0.035 | 0.032 | 0.053 | −0.071 | −0.250 |
| 275.151 | 10.02315 | 98.095 | 0.034 | 0.035 | 0.043 | −0.024 | −0.217 |
| 275.151 | 9.01456 | 86.332 | 0.033 | 0.038 | 0.030 | 0.018 | −0.182 |
| 275.149 | 8.01243 | 74.910 | 0.031 | 0.042 | 0.008 | 0.044 | −0.156 |
| 275.148 | 7.00879 | 63.851 | 0.030 | 0.047 | −0.013 | 0.058 | −0.132 |
| 275.148 | 6.00792 | 53.274 | 0.029 | 0.054 | −0.036 | 0.058 | −0.116 |
| 275.148 | 5.00582 | 43.180 | 0.028 | 0.064 | −0.052 | 0.052 | −0.099 |
| 275.150 | 4.00504 | 33.606 | 0.027 | 0.079 | −0.055 | 0.046 | −0.077 |
| 275.153 | 2.99096 | 24.415 | 0.026 | 0.105 | −0.030 | 0.058 | −0.035 |
| 275.150 | 2.00416 | 15.931 | 0.025 | 0.155 | −0.021 | 0.046 | −0.016 |
| 275.153 | 1.00349 | 7.771 | 0.024 | 0.306 | 0.017 | 0.056 | 0.025 |
| | | | | 300.000 K | | | |
| 300.098 | 19.86189 | 167.813 | 0.042 | 0.025 | −0.072 | −0.220 | −0.166 |
| 300.097 | 19.04526 | 161.669 | 0.041 | 0.026 | −0.067 | −0.221 | −0.188 |
| 300.098 | 18.02223 | 153.661 | 0.040 | 0.026 | −0.059 | −0.215 | −0.207 |
| 300.095 | 17.02152 | 145.498 | 0.039 | 0.027 | −0.056 | −0.209 | −0.222 |
| 300.095 | 16.02262 | 137.045 | 0.039 | 0.028 | −0.051 | −0.197 | −0.228 |
| 300.096 | 15.02094 | 128.298 | 0.038 | 0.029 | −0.045 | −0.179 | −0.226 |
| 300.096 | 14.02039 | 119.331 | 0.036 | 0.031 | −0.040 | −0.158 | −0.219 |
| 300.097 | 13.01449 | 110.137 | 0.035 | 0.032 | −0.037 | −0.135 | −0.209 |
| 300.097 | 12.01914 | 100.924 | 0.034 | 0.034 | −0.037 | −0.113 | −0.197 |
| 300.097 | 11.01575 | 91.591 | 0.033 | 0.036 | −0.035 | −0.085 | −0.179 |
| 300.098 | 10.01397 | 82.283 | 0.032 | 0.039 | −0.044 | −0.068 | −0.170 |
| 300.099 | 9.01304 | 73.068 | 0.031 | 0.043 | −0.051 | −0.050 | −0.157 |
| 300.096 | 8.00923 | 63.965 | 0.030 | 0.047 | −0.061 | −0.038 | −0.146 |
| 300.096 | 7.00832 | 55.076 | 0.029 | 0.053 | −0.067 | −0.027 | −0.133 |
| 300.100 | 6.00560 | 46.394 | 0.028 | 0.061 | −0.072 | −0.021 | −0.119 |
| 300.100 | 5.00492 | 37.982 | 0.027 | 0.072 | −0.076 | −0.020 | −0.107 |
| 300.099 | 4.00431 | 29.837 | 0.026 | 0.088 | −0.078 | −0.022 | −0.095 |
| 300.098 | 2.98941 | 21.862 | 0.025 | 0.116 | −0.060 | −0.010 | −0.066 |
| 300.099 | 2.00304 | 14.382 | 0.025 | 0.171 | −0.045 | −0.005 | −0.044 |
| 300.103 | 1.00294 | 7.070 | 0.024 | 0.335 | 0.001 | 0.025 | 0.005 |
| | | | | 325.000 K | | | |
| 325.097 | 19.86308 | 145.342 | 0.039 | 0.027 | −0.014 | −0.109 | −0.094 |
| 325.097 | 19.04171 | 139.754 | 0.039 | 0.028 | −0.015 | −0.107 | −0.099 |
| 325.097 | 18.04301 | 132.773 | 0.038 | 0.029 | −0.020 | −0.107 | −0.106 |
| 325.095 | 17.03409 | 125.540 | 0.037 | 0.030 | −0.018 | −0.100 | −0.105 |
| 325.095 | 16.03472 | 118.203 | 0.036 | 0.031 | −0.018 | −0.092 | −0.103 |
| 325.095 | 15.01273 | 110.548 | 0.035 | 0.032 | −0.017 | −0.083 | −0.100 |
| 325.093 | 14.02712 | 103.047 | 0.035 | 0.034 | −0.017 | −0.073 | −0.097 |
| 325.093 | 13.02177 | 95.305 | 0.034 | 0.035 | −0.019 | −0.062 | −0.094 |
| 325.093 | 12.02022 | 87.533 | 0.033 | 0.038 | −0.021 | −0.052 | −0.091 |
| 325.093 | 11.01652 | 79.725 | 0.032 | 0.040 | −0.017 | −0.035 | −0.081 |
| 325.094 | 10.01210 | 71.912 | 0.031 | 0.043 | −0.023 | −0.028 | −0.080 |
| 325.093 | 9.00774 | 64.144 | 0.030 | 0.047 | −0.025 | −0.019 | −0.075 |
| 325.095 | 8.00864 | 56.486 | 0.029 | 0.052 | −0.028 | −0.012 | −0.070 |
| 325.095 | 7.00809 | 48.915 | 0.028 | 0.058 | −0.029 | −0.004 | −0.062 |
| 325.095 | 6.00585 | 41.448 | 0.028 | 0.067 | −0.032 | −0.002 | −0.057 |
| 325.095 | 5.00480 | 34.129 | 0.027 | 0.078 | −0.036 | −0.003 | −0.052 |
| 325.096 | 4.00386 | 26.964 | 0.026 | 0.096 | −0.039 | −0.007 | −0.049 |
| 325.099 | 2.98777 | 19.857 | 0.025 | 0.127 | −0.045 | −0.016 | −0.049 |
| 325.099 | 2.00334 | 13.144 | 0.024 | 0.186 | −0.027 | −0.003 | −0.027 |
| 325.102 | 1.00308 | 6.494 | 0.024 | 0.364 | 0.001 | 0.016 | 0.004 |
| | | | | 350.000 K | | | |
| 350.086 | 19.73559 | 127.910 | 0.037 | 0.029 | −0.027 | −0.074 | −0.053 |
| 350.086 | 19.03490 | 123.623 | 0.037 | 0.030 | −0.031 | −0.074 | −0.055 |
| 350.086 | 18.01221 | 117.253 | 0.036 | 0.031 | −0.035 | −0.073 | −0.057 |
| 350.087 | 17.02173 | 110.962 | 0.036 | 0.032 | −0.038 | −0.072 | −0.058 |
| 350.088 | 16.01177 | 104.441 | 0.035 | 0.033 | −0.039 | −0.067 | −0.058 |
| 350.089 | 15.01162 | 97.891 | 0.034 | 0.035 | −0.039 | −0.062 | −0.058 |
| 350.087 | 14.01032 | 91.255 | 0.033 | 0.036 | −0.040 | −0.058 | −0.058 |
| 350.088 | 13.01344 | 84.590 | 0.033 | 0.038 | −0.039 | −0.051 | −0.057 |
| 350.087 | 12.00734 | 77.821 | 0.032 | 0.041 | −0.038 | −0.045 | −0.057 |
| 350.087 | 11.00656 | 71.069 | 0.031 | 0.044 | −0.033 | −0.034 | −0.051 |
| 350.088 | 10.00614 | 64.310 | 0.030 | 0.047 | −0.035 | −0.031 | −0.052 |
| 350.085 | 9.00469 | 57.562 | 0.029 | 0.051 | −0.035 | −0.025 | −0.051 |
| 350.087 | 8.00657 | 50.867 | 0.029 | 0.056 | −0.036 | −0.022 | −0.050 |
| 350.086 | 7.00340 | 44.191 | 0.028 | 0.063 | −0.032 | −0.016 | −0.044 |
| 350.095 | 6.00561 | 37.612 | 0.027 | 0.072 | −0.032 | −0.013 | −0.041 |
| 350.097 | 5.00391 | 31.086 | 0.026 | 0.085 | −0.034 | −0.013 | −0.039 |
| 350.096 | 4.00402 | 24.661 | 0.026 | 0.104 | −0.038 | −0.017 | −0.039 |
| 350.092 | 2.98364 | 18.205 | 0.025 | 0.137 | −0.054 | −0.035 | −0.052 |
| 350.093 | 2.00254 | 12.108 | 0.024 | 0.200 | −0.043 | −0.027 | −0.040 |
| 350.086 | 1.00257 | 6.004 | 0.024 | 0.393 | −0.025 | −0.014 | −0.021 |

[(a)] Expanded uncertainties ($k$ = 2): $U(p > 3)/\mathrm{MPa} = 7.5 \cdot 10^{-5} \cdot \frac{p}{\mathrm{MPa}} + 4 \cdot 10^{-3}$; $U(p < 3)/\mathrm{MPa} = 6.0 \cdot 10^{-5} \cdot \frac{p}{\mathrm{MPa}} + 2 \cdot 10^{-3}$; $U(T)$ = 15 mK; $\frac{U(\rho)}{\mathrm{kg \cdot m^{-3}}} = 2.5 \cdot 10^{4} \frac{\chi_S}{m^3 kg^{-1}} + 1.1 \cdot 10^{-4} \cdot \frac{\rho}{\mathrm{kg \cdot m^{-3}}} + 2.3 \cdot 10^{-2}$.

**Table 5.** Experimental ($p$, $\rho_{exp}$, $T$) measurements for the ($H_2$-enriched) NG mixture G 456 (20 % $H_2$), absolute and relative expanded ($k$ = 2) uncertainty in density, $U(\rho_{exp})$, relative deviations from the density given by the AGA8-DC92 EoS [19], $\rho_{AGA8\text{-}DC92}$, the GERG-2008 EoS [20,21], $\rho_{GERG\text{-}2008}$, and the improved GERG-2008 EoS [52–54], $\rho_{GERG\text{-}improved}$.

| $T$ / K[(a)] | $p$ / MPa[(a)] | $\rho_{exp}$ / kg·m$^{-3}$[(a)] | $U(\rho_{exp})$ / kg·m$^{-3}$ | $10^2\ U(\rho_{exp})/\rho_{exp}$ | $10^2\ (\rho_{exp} - \rho_{AGA8\text{-}DC92})/\rho_{AGA8\text{-}DC92}$ | $10^2\ (\rho_{exp} - \rho_{GERG\text{-}2008})/\rho_{GERG\text{-}2008}$ | $10^2\ (\rho_{exp} - \rho_{GERG\text{-}improved})/\rho_{GERG\text{-}improved}$ |
|---|---|---|---|---|---|---|---|
| | | | | 260.000 K | | | |
| 260.174 | 19.60635 | 180.852 | 0.043 | 0.024 | −0.041 | −0.343 | −0.108 |
| 260.176 | 19.06388 | 176.807 | 0.043 | 0.024 | −0.006 | −0.334 | −0.145 |
| 260.176 | 18.08779 | 169.185 | 0.042 | 0.025 | 0.051 | −0.313 | −0.210 |
| 260.177 | 17.08145 | 160.855 | 0.041 | 0.026 | 0.108 | −0.279 | −0.262 |
| 260.176 | 16.07121 | 152.011 | 0.040 | 0.026 | 0.161 | −0.230 | −0.296 |
| 260.174 | 15.05336 | 142.618 | 0.039 | 0.027 | 0.206 | −0.166 | −0.305 |
| 260.176 | 14.04961 | 132.913 | 0.038 | 0.029 | 0.244 | −0.089 | −0.288 |
| 260.177 | 13.04327 | 122.804 | 0.037 | 0.030 | 0.271 | −0.001 | −0.248 |
| 260.176 | 12.03722 | 112.398 | 0.036 | 0.032 | 0.284 | 0.087 | −0.195 |
| 260.174 | 11.02706 | 101.759 | 0.034 | 0.034 | 0.284 | 0.173 | −0.134 |
| 260.177 | 10.02106 | 91.082 | 0.033 | 0.037 | 0.261 | 0.237 | −0.086 |
| 260.177 | 9.01720 | 80.487 | 0.032 | 0.040 | 0.221 | 0.278 | −0.050 |
| 260.178 | 8.01319 | 70.064 | 0.031 | 0.044 | 0.169 | 0.293 | −0.030 |
| 260.178 | 7.01101 | 59.936 | 0.030 | 0.050 | 0.116 | 0.286 | −0.019 |
| 260.176 | 6.00829 | 50.148 | 0.029 | 0.057 | 0.059 | 0.255 | −0.023 |
| 260.177 | 5.00601 | 40.759 | 0.028 | 0.068 | 0.012 | 0.212 | −0.028 |
| 260.177 | 4.00470 | 31.797 | 0.027 | 0.083 | −0.020 | 0.166 | −0.030 |
| 260.174 | 2.99387 | 23.178 | 0.026 | 0.110 | −0.020 | 0.137 | −0.011 |
| 260.172 | 2.00318 | 15.134 | 0.025 | 0.163 | 0.003 | 0.119 | 0.021 |
| 260.178 | 1.00301 | 7.400 | 0.024 | 0.321 | 0.095 | 0.158 | 0.109 |
| | | | | 275.000 K | | | |
| 275.160 | 19.45538 | 161.569 | 0.041 | 0.026 | −0.031 | −0.255 | −0.206 |
| 275.160 | 18.04039 | 151.090 | 0.040 | 0.027 | 0.010 | −0.228 | −0.246 |
| 275.156 | 17.02701 | 143.171 | 0.039 | 0.027 | 0.033 | −0.204 | −0.262 |
| 275.158 | 16.04189 | 135.157 | 0.038 | 0.028 | 0.057 | −0.169 | −0.262 |
| 275.156 | 15.03640 | 126.678 | 0.037 | 0.029 | 0.075 | −0.130 | −0.252 |
| 275.158 | 14.03271 | 117.949 | 0.036 | 0.031 | 0.090 | −0.085 | −0.230 |
| 275.157 | 13.03333 | 109.043 | 0.035 | 0.032 | 0.095 | −0.041 | −0.206 |
| 275.156 | 12.02664 | 99.915 | 0.034 | 0.034 | 0.093 | 0.002 | −0.180 |
| 275.158 | 11.02115 | 90.714 | 0.033 | 0.037 | 0.090 | 0.049 | −0.149 |
| 275.157 | 10.01815 | 81.512 | 0.032 | 0.039 | 0.070 | 0.077 | −0.131 |
| 275.157 | 9.01514 | 72.368 | 0.031 | 0.043 | 0.047 | 0.099 | −0.115 |
| 275.159 | 8.01265 | 63.348 | 0.030 | 0.048 | 0.020 | 0.110 | −0.103 |
| 275.155 | 7.00790 | 54.488 | 0.029 | 0.053 | −0.006 | 0.111 | −0.095 |
| 275.155 | 6.00636 | 45.879 | 0.028 | 0.061 | −0.028 | 0.105 | −0.085 |
| 275.155 | 5.00625 | 37.537 | 0.027 | 0.072 | −0.044 | 0.092 | −0.074 |
| 275.153 | 4.00470 | 29.459 | 0.026 | 0.089 | −0.045 | 0.083 | −0.055 |
| 275.155 | 2.99059 | 21.574 | 0.025 | 0.117 | −0.025 | 0.084 | −0.021 |
| 275.155 | 2.00369 | 14.179 | 0.024 | 0.173 | −0.005 | 0.076 | 0.005 |
| 275.159 | 1.00341 | 6.964 | 0.024 | 0.340 | 0.038 | 0.084 | 0.048 |
| | | | | 300.000 K | | | |
| 300.107 | 19.87742 | 141.339 | 0.039 | 0.028 | 0.053 | −0.075 | −0.062 |
| 300.108 | 19.02634 | 135.801 | 0.038 | 0.028 | 0.061 | −0.065 | −0.064 |
| 300.109 | 18.02164 | 129.075 | 0.038 | 0.029 | 0.070 | −0.052 | −0.063 |
| 300.110 | 17.02716 | 122.225 | 0.037 | 0.030 | 0.077 | −0.036 | −0.057 |
| 300.109 | 16.02478 | 115.143 | 0.036 | 0.031 | 0.083 | −0.019 | −0.051 |
| 300.109 | 15.01494 | 107.849 | 0.035 | 0.033 | 0.087 | −0.002 | −0.043 |
| 300.105 | 14.01853 | 100.521 | 0.034 | 0.034 | 0.087 | 0.016 | −0.036 |
| 300.107 | 13.01552 | 93.038 | 0.033 | 0.036 | 0.086 | 0.034 | −0.029 |
| 300.107 | 12.01365 | 85.490 | 0.033 | 0.038 | 0.080 | 0.050 | −0.025 |
| 300.106 | 11.01219 | 77.912 | 0.032 | 0.041 | 0.078 | 0.071 | −0.016 |
| 300.104 | 10.01067 | 70.322 | 0.031 | 0.044 | 0.067 | 0.081 | −0.015 |
| 300.107 | 9.00988 | 62.766 | 0.030 | 0.048 | 0.057 | 0.091 | −0.013 |
| 300.106 | 8.00698 | 55.254 | 0.029 | 0.053 | 0.046 | 0.097 | −0.010 |
| 300.107 | 7.00700 | 47.853 | 0.028 | 0.059 | 0.038 | 0.102 | −0.004 |
| 300.108 | 6.00601 | 40.556 | 0.028 | 0.068 | 0.030 | 0.101 | 0.001 |
| 300.107 | 5.00426 | 33.388 | 0.027 | 0.080 | 0.023 | 0.096 | 0.006 |
| 300.109 | 4.00415 | 26.380 | 0.026 | 0.098 | 0.022 | 0.091 | 0.016 |
| 300.108 | 3.00315 | 19.528 | 0.025 | 0.129 | 0.023 | 0.083 | 0.025 |
| 300.106 | 2.00269 | 12.849 | 0.024 | 0.189 | 0.033 | 0.079 | 0.040 |
| 300.114 | 1.00270 | 6.346 | 0.024 | 0.372 | 0.063 | 0.090 | 0.070 |
| | | | | 325.000 K | | | |
| 325.102 | 19.92240 | 124.409 | 0.037 | 0.030 | −0.038 | −0.107 | −0.066 |
| 325.102 | 19.03473 | 119.239 | 0.036 | 0.031 | −0.037 | −0.102 | −0.065 |
| 325.104 | 18.02485 | 113.235 | 0.036 | 0.032 | −0.034 | −0.093 | −0.060 |
| 325.101 | 17.02659 | 107.176 | 0.035 | 0.033 | −0.033 | −0.086 | −0.059 |
| 325.104 | 16.02497 | 100.987 | 0.034 | 0.034 | −0.030 | −0.075 | −0.055 |
| 325.105 | 15.01994 | 94.679 | 0.034 | 0.036 | −0.028 | −0.065 | −0.052 |
| 325.105 | 14.01639 | 88.297 | 0.033 | 0.037 | −0.027 | −0.055 | −0.051 |
| 325.105 | 13.01781 | 81.881 | 0.032 | 0.039 | −0.027 | −0.045 | −0.051 |
| 325.104 | 12.01602 | 75.396 | 0.031 | 0.042 | −0.027 | −0.034 | −0.051 |
| 325.104 | 11.01593 | 68.894 | 0.031 | 0.045 | −0.024 | −0.021 | −0.048 |
| 325.104 | 10.01188 | 62.350 | 0.030 | 0.048 | −0.029 | −0.016 | −0.051 |
| 325.103 | 9.00880 | 55.821 | 0.029 | 0.052 | −0.031 | −0.009 | −0.052 |
| 325.102 | 8.00910 | 49.339 | 0.029 | 0.058 | −0.034 | −0.005 | −0.052 |
| 325.103 | 7.00594 | 42.877 | 0.028 | 0.065 | −0.035 | $< 0.001$ | −0.049 |
| 325.102 | 6.00544 | 36.488 | 0.027 | 0.074 | −0.041 | −0.002 | −0.050 |
| 325.104 | 5.00397 | 30.164 | 0.026 | 0.087 | −0.046 | −0.006 | −0.050 |
| 325.104 | 4.00403 | 23.933 | 0.026 | 0.107 | −0.050 | −0.012 | −0.049 |
| 325.105 | 2.98910 | 17.703 | 0.025 | 0.141 | −0.065 | −0.031 | −0.060 |
| 325.104 | 2.00279 | 11.755 | 0.024 | 0.206 | −0.053 | −0.026 | −0.046 |
| 325.110 | 1.00294 | 5.831 | 0.024 | 0.404 | −0.030 | −0.013 | −0.024 |
| | | | | 350.000 K | | | |
| 350.091 | 19.94877 | 111.610 | 0.036 | 0.032 | −0.050 | −0.080 | −0.013 |
| 350.093 | 19.03368 | 106.810 | 0.035 | 0.033 | −0.049 | −0.076 | −0.012 |
| 350.094 | 18.02103 | 101.408 | 0.034 | 0.034 | −0.049 | −0.071 | −0.011 |
| 350.092 | 17.01185 | 95.937 | 0.034 | 0.035 | −0.049 | −0.068 | −0.012 |
| 350.092 | 16.01717 | 90.469 | 0.033 | 0.037 | −0.047 | −0.062 | −0.012 |
| 350.092 | 15.00954 | 84.858 | 0.033 | 0.038 | −0.046 | −0.056 | −0.014 |
| 350.091 | 14.00976 | 79.231 | 0.032 | 0.040 | −0.044 | −0.049 | −0.016 |
| 350.092 | 13.00725 | 73.537 | 0.031 | 0.043 | −0.042 | −0.043 | −0.019 |
| 350.091 | 12.00753 | 67.820 | 0.031 | 0.045 | −0.041 | −0.038 | −0.022 |
| 350.092 | 11.00601 | 62.068 | 0.030 | 0.048 | −0.034 | −0.026 | −0.019 |
| 350.092 | 10.00977 | 56.325 | 0.029 | 0.052 | −0.035 | −0.023 | −0.023 |
| 350.092 | 9.00567 | 50.531 | 0.029 | 0.057 | −0.034 | −0.019 | −0.025 |
| 350.092 | 8.00758 | 44.777 | 0.028 | 0.063 | −0.036 | −0.017 | −0.029 |
| 350.089 | 7.00517 | 39.016 | 0.027 | 0.070 | −0.036 | −0.015 | −0.029 |
| 350.090 | 6.00418 | 33.287 | 0.027 | 0.080 | −0.043 | −0.020 | −0.036 |
| 350.093 | 5.00418 | 27.602 | 0.026 | 0.094 | −0.049 | −0.026 | −0.041 |
| 350.093 | 4.00362 | 21.959 | 0.025 | 0.116 | −0.060 | −0.038 | −0.051 |
| 350.091 | 2.98613 | 16.278 | 0.025 | 0.152 | −0.071 | −0.050 | −0.060 |
| 350.093 | 2.00296 | 10.848 | 0.024 | 0.222 | −0.089 | −0.072 | −0.079 |
| 350.100 | 1.00296 | 5.395 | 0.023 | 0.435 | −0.091 | −0.080 | −0.084 |

[a] Expanded uncertainties ($k$ = 2): $U(p > 3)/\mathrm{MPa} = 7.5 \cdot 10^{-5} \cdot \frac{p}{\mathrm{MPa}} + 4 \cdot 10^{-3}$; $U(p < 3)/\mathrm{MPa} = 6.0 \cdot 10^{-5} \cdot \frac{p}{\mathrm{MPa}} + 2 \cdot 10^{-3}$; $U(T)$ = 15 mK; $\frac{U(\rho)}{\mathrm{kg \cdot m^{-3}}} = 2.5 \cdot 10^{4} \frac{\chi_S}{m^3 kg^{-1}} + 1.1 \cdot 10^{-4} \cdot \frac{\rho}{\mathrm{kg \cdot m^{-3}}} + 2.3 \cdot 10^{-2}$.

**Table 6.** Statistical analysis of the ($p$, $\rho$, $T$) data sets in relation to the AGA8-DC92 EoS [19], GERG-2008 EoS [20,21], and improved GERG-2008 EoS [52–54] for all the NG mixtures investigated in this study, including literature data for similar mixtures. *AARD* = average absolute value of the relative deviations, *BiasRD* = average relative deviation, *RMSRD* = root mean square relative deviation, *MaxRD* = maximum value of the relative deviations.

| Reference[(a)] | $x_{H_2}$ | $N$[(b)] | Covered ranges | | Experimental *vs* AGA8-DC92 EoS | | | | Experimental *vs* GERG-2008 EoS | | | | Experimental *vs* improved GERG-2008 EoS | | | |
|---|---|---|---|---|---|---|---|---|---|---|---|---|---|---|---|---|
| | | | $T$ / K | $p$ / MPa | *AARD* / % | *BiasRD* / % | *RMSRD* / % | *MaxRD* / % | *AARD* / % | *BiasRD* / % | *RMSRD* / % | *MaxRD* / % | *AARD* / % | *BiasRD* / % | *RMSRD* / % | *MaxRD* / % |
| G 432 (this work) | 0 | 93 | 260–350 | 1–20 | 0.040 | −0.033 | 0.055 | 0.15 | 0.043 | −0.023 | 0.063 | 0.21 | 0.062 | −0.051 | 0.088 | 0.28 |
| G 455 (this work) | 0.099864 | 98 | 260–350 | 1–20 | 0.048 | −0.016 | 0.062 | 0.19 | 0.11 | −0.074 | 0.16 | 0.44 | 0.12 | −0.12 | 0.15 | 0.38 |
| G 456 (this work) | 0.200039 | 99 | 260–350 | 1–20 | 0.065 | 0.025 | 0.088 | 0.28 | 0.094 | −0.0039 | 0.12 | 0.34 | 0.077 | −0.070 | 0.11 | 0.30 |
| Lozano-Martín et al., 2024 [32] | 0 | 97 | 250–350 | 1–20 | 0.012 | −0.0066 | 0.018 | 0.054 | 0.032 | 0.032 | 0.034 | 0.049 | 0.019 | 0.016 | 0.023 | 0.051 |
| Lozano-Martín et al., 2024 [32] | 0.099928 | 98 | 275–350 | 1–20 | 0.032 | −0.029 | 0.045 | 0.19 | 0.029 | −0.021 | 0.049 | 0.18 | 0.047 | −0.042 | 0.070 | 0.20 |
| Lozano-Martín et al., 2024 [32] | 0.199945 | 99 | 250–350 | 1–20 | 0.033 | −0.020 | 0.039 | 0.12 | 0.030 | −0.015 | 0.037 | 0.11 | 0.052 | −0.043 | 0.072 | 0.19 |

[(a)] Only vapor and supercritical phase measurements have been considered.

[(b)] Number of experimental points.

**Table 7.** Derived isothermal compressibility $\kappa_T$ values for all the reference NG mixtures examined in this study at various temperatures $T$ and pressures $p$.

| | $\kappa_T$ / MPa$^{-1}$ [(a)] | | | | |
|---|---|---|---|---|---|
| | $T$ / K | | | | |
| $p$ / MPa | 260 | 275 | 300 | 325 | 350 |
| | | | G 432 | | |
| 19 | | | | 0.0484 | 0.0496 |
| 18 | | | 0.0494 | 0.0532 | 0.0542 |
| 17 | 0.0361 | 0.0460 | 0.0556 | 0.0583 | 0.0586 |
| 16 | 0.0418 | 0.0535 | 0.0623 | 0.0639 | 0.0634 |
| 15 | 0.0512 | 0.0625 | 0.0698 | 0.0700 | 0.0688 |
| 14 | 0.0621 | 0.0732 | 0.0780 | 0.0768 | 0.0750 |
| 13 | 0.0763 | 0.0859 | 0.0871 | 0.0843 | 0.0817 |
| 12 | 0.0942 | 0.0999 | 0.0970 | 0.0925 | 0.0893 |
| 11 | 0.1155 | 0.1152 | 0.1078 | 0.1019 | 0.0980 |
| 10 | 0.1386 | 0.1311 | 0.1196 | 0.1126 | 0.1082 |
| 9 | 0.1610 | 0.1473 | 0.1329 | 0.1251 | 0.1202 |
| 8 | 0.1816 | 0.1644 | 0.1484 | 0.1401 | 0.1350 |
| 7 | 0.2017 | 0.1836 | 0.1674 | 0.1588 | 0.1536 |
| 6 | 0.2243 | 0.2075 | 0.1919 | 0.1834 | 0.1781 |
| 5 | 0.2546 | 0.2399 | 0.2254 | 0.2172 | 0.2119 |
| 4 | 0.3008 | 0.2884 | 0.2753 | 0.2675 | 0.2621 |
| 3 | 0.3802 | 0.3702 | 0.3587 | 0.3514 | 0.3463 |
| 2 | 0.5427 | 0.5311 | 0.5206 | 0.5135 | 0.5031 |
| | | | G 455 (G 432 + 10 % $H_2$) | | |
| 19 | | 0.0422 | 0.0474 | 0.0492 | 0.0497 |
| 18 | | 0.0474 | 0.0522 | 0.0535 | 0.0537 |
| 17 | 0.0478 | 0.0532 | 0.0573 | 0.0580 | 0.0578 |
| 16 | 0.0548 | 0.0599 | 0.0629 | 0.0629 | 0.0623 |
| 15 | 0.0632 | 0.0675 | 0.0692 | 0.0684 | 0.0674 |
| 14 | 0.0730 | 0.0761 | 0.0761 | 0.0745 | 0.0730 |
| 13 | 0.0844 | 0.0856 | 0.0838 | 0.0813 | 0.0794 |
| 12 | 0.0972 | 0.0961 | 0.0922 | 0.0890 | 0.0867 |
| 11 | 0.1112 | 0.1076 | 0.1018 | 0.0978 | 0.0951 |
| 10 | 0.1263 | 0.1202 | 0.1127 | 0.1081 | 0.1050 |
| 9 | 0.1422 | 0.1342 | 0.1254 | 0.1202 | 0.1169 |
| 8 | 0.1595 | 0.1502 | 0.1406 | 0.1351 | 0.1315 |
| 7 | 0.1793 | 0.1695 | 0.1595 | 0.1538 | 0.1500 |
| 6 | 0.2038 | 0.1942 | 0.1841 | 0.1783 | 0.1744 |
| 5 | 0.2368 | 0.2278 | 0.2181 | 0.2123 | 0.2083 |
| 4 | 0.2859 | 0.2776 | 0.2684 | 0.2628 | 0.2587 |
| 3 | 0.3678 | 0.3607 | 0.3520 | 0.3467 | 0.3429 |
| 2 | 0.5330 | 0.5243 | 0.5155 | 0.5100 | 0.5014 |
| | | | G 456 (G 432 + 20 % $H_2$) | | |
| 19 | 0.0433 | 0.0462 | 0.0484 | 0.0492 | 0.0492 |
| 18 | 0.0480 | 0.0505 | 0.0528 | 0.0532 | 0.0532 |
| 17 | 0.0535 | 0.0559 | 0.0572 | 0.0573 | 0.0570 |
| 16 | 0.0597 | 0.0615 | 0.0622 | 0.0618 | 0.0613 |
| 15 | 0.0668 | 0.0679 | 0.0677 | 0.0669 | 0.0660 |
| 14 | 0.0747 | 0.0749 | 0.0739 | 0.0725 | 0.0714 |
| 13 | 0.0836 | 0.0828 | 0.0808 | 0.0789 | 0.0776 |
| 12 | 0.0934 | 0.0915 | 0.0885 | 0.0862 | 0.0846 |
| 11 | 0.1044 | 0.1014 | 0.0974 | 0.0947 | 0.0928 |
| 10 | 0.1167 | 0.1126 | 0.1078 | 0.1047 | 0.1025 |
| 9 | 0.1305 | 0.1256 | 0.1200 | 0.1166 | 0.1142 |
| 8 | 0.1466 | 0.1411 | 0.1350 | 0.1312 | 0.1287 |
| 7 | 0.1661 | 0.1602 | 0.1537 | 0.1498 | 0.1471 |
| 6 | 0.1910 | 0.1850 | 0.1784 | 0.1743 | 0.1715 |
| 5 | 0.2249 | 0.2190 | 0.2123 | 0.2082 | 0.2053 |
| 4 | 0.2749 | 0.2693 | 0.2628 | 0.2587 | 0.2557 |
| 3 | 0.3578 | 0.3530 | 0.3468 | 0.3427 | 0.3399 |
| 2 | 0.5250 | 0.5181 | 0.5113 | 0.5069 | 0.4998 |

(a) Expanded uncertainty ($k = 2$): $U_r(\kappa_T) = 0.7$ %.